\documentclass[10pt,journal,compsoc]{IEEEtran}
\usepackage{filecontents}
\usepackage{amsmath,amsthm,dsfont,tabularx,scalefnt,comment}
\usepackage[cmintegrals]{newtxmath}
\usepackage[nocompress]{cite}
\usepackage{subcaption}
\usepackage{algorithm,algorithmic}
\usepackage{caption}
\usepackage{bm}
\usepackage[noblocks]{authblk}
\usepackage{color,soul}
\usepackage{multirow}
\usepackage{siunitx}
\usepackage{setspace,adjustbox}

\newtheorem{theorem}{Theorem}[section]

\floatname{Procedure}{algorithm}

\newtheorem{definition}{Definition}[section]

\newtheorem{lemma}[theorem]{Lemma}

\begin{document}
\title{Two Birds with One Stone: Differential Privacy by Low-power SRAM Memory}
\author{Jianqing Liu*,~\IEEEmembership{Member,~IEEE}
        Na Gong*,~\IEEEmembership{Member,~IEEE}
        Hritom Das,~\IEEEmembership{Member,~IEEE}

\thanks{
* co-primary authors.}
\thanks{
J. Liu is with the Department
of Computer Science, North Carolina State University, Raleigh,
NC 27606 USA e-mail: jliu96@ncsu.edu}
\thanks{N. Gong is with the Department of Electrical and Computer Engineering, University of South Alabama, Mobile, AL 36688 USA email: nagong@southalabama.edu}
\thanks{H. Das is with the Department of Electrical and Computer Engineering, University of Tennessee, Knoxville, TN 37996 USA e-mail: hdas@utk.edu}
\thanks{The work by J. Liu was supported in part by the National Science Foundation under grants ECCS-2312738 and CNS-2247273. The work by N. Gong was supported in part by the National Science Foundation under grants CNS-2211215 and OIA-2218046.}
}

\IEEEtitleabstractindextext{
\begin{abstract}
The software-based implementation of differential privacy mechanisms has been shown to be neither friendly for lightweight devices nor secure against side-channel attacks. In this work, we aim to develop a hardware-based technique to achieve differential privacy by design. In contrary to the conventional software-based noise generation and injection process, our design realizes local differential privacy (LDP) by harnessing the inherent hardware noise into controlled LDP noise when data is stored in the memory. Specifically, the noise is tamed through a novel memory design and power downscaling technique, which leads to double-faceted gains in privacy and power efficiency. A well-round study that consists of theoretical design and analysis and chip implementation and experiments is presented. The results confirm that the developed technique is differentially private, saves 88.58\% system power, speeds up software-based DP mechanisms by more than $10^6$ times, while only incurring 2.46\% chip overhead and 7.81\% estimation errors in data recovery. 
\end{abstract}
}

\maketitle

\section{Introduction}
Nowadays, with the rapid development of Internet-of-Things (IoT) technologies, the collection of user's and environment's data becomes immense. Such a rich set of data are frequently aggregated and analyzed using advanced Artificial Intelligence (AI) algorithms to make faster and more informed decisions. Despite the increased convenience and intelligence, the collection and analysis of data could incur significant privacy risks which are evidenced by recent privacy breach incidents such as historical data collected from Netflix \cite{zhou2008large} leading to unwanted intrusive marketing. Preserving data privacy thus attracts unprecedented amount of attention in this era.

The line of work on data privacy has a groundbreaking advance with the seminal work by Dwork et al. \cite{dwork2006calibrating} in differential privacy (DP), which now becomes the \emph{de facto} standard for preserving data privacy. The essence of the DP notion is to ensure the rigourous privacy guarantee of individual's data without sacrificing its general statistics. Local differential privacy (LDP), in contrast to the traditional DP notion in the centralized setting \cite{machanavajjhala2008privacy,mcsherry2009differentially}, is the state-of-the-art approach which perturbs data locally with guarantees of plausible deniability \cite{qin2016heavy}. As such, LDP protects individual's data against the untrusted curator and still preserves useful statistics of the original data. One example of the LDP realization is RAPPOR which was developed by Google and is currently implemented in its Chrome browser \cite{erlingsson2014rappor}.

However, existing realization or implementation of DP (including LDP) have many issues. That is, existing DP mechanisms typically add (subtract) to (from) the accurate value a secret number (noise) sampled from a probability distribution, which unfortunately bears the following drawbacks. (i) The sampling/adding/subtrating process in DP mechanisms is converted to the floating-point arithmetic in the operating system (OS), which actually deviates markedly from the DP's mathematical abstraction, due to rounding rules, compounding errors and many more in the floating-point arithmetic, and could lead to side-channel attacks. The vulnerability was spotted by Mironov et al. \cite{mironov2012significance} and later exploited by Andrysco et al. \cite{andrysco2015subnormal} to launch attacks to the Fuzz differentially private database \cite{haeberlen2011differential}. (ii) The noise sampling process in DP mechanisms requires function calls in the high-layer protocol stack, which are easily supported by legacy OSs such as iOS, Linux and Windows but may not stand for ``slim'' OSs in lightweight IoT devices such as sensors, cameras or lightbulbs. These IoT OSs are mostly vendor- and application-specific, which makes the implementation of DP mechanisms difficult to scale. (iii) The computation processes of DP mechanisms involve a significant number of CPU calls and memory accesses, which could be resource-consuming and power hungry. Although the cost of one-time DP randomization could be negligible, it will be prohibitively high for real-time or steaming data collection services when computations repeat. The consumed resources and energy consumption could become prohibitive for lightweight IoT devices such as streaming cameras and wearable ECG sensors.

The above concerns thus call for a more secure, scalable, and lightweight implementation of DP mechanisms, but very limited number of works have set foot on this research problem. To fill this gap, this paper develops a novel hardware-based technique that perturbs data when it is stored in memory --- an ubiquitous component in all electronic devices --- thereby accomplishing the vision of (differential) \emph{privacy by design} \cite{liu2023privacy}. In contrast to existing hardware security primitives like physical unclonable function (PUF), we do not seek to generate and later use hardware noises but rather perturb the data in situ such that it carries DP noises naturally. Moreover, compared with other hardware components like CPU, achieving DP in memory avoids data passing between CPU and memories. This design idea of computation-storage integration is reminiscent of a new concept --- ``computing in memory'' (CIM) which is inspired by the highly energy-efficient mammalian brain where memory and processing are deeply intertwined \cite{CIP1}. In addition, by minimizing system interactions, we can also avoid threat interfaces and reduce system overhead.

To this end, we develop a novel static random access memory (SRAM) architecture, coined as \texttt{SRAM\_DP}, to achieve \emph{randomized response (RR)}  \cite{warner1965randomized} --- the widely adopted approach to achieve LDP. The novelty of \texttt{SRAM\_DP} lies in its integration of data storage and LDP and enhanced power efficiency. Specifically, by reducing the memory's supply voltage, each custom memory cell becomes volatile and has a probability $f$ to fail thus flipping the stored bit ($1 \rightarrow 0$ or $0 \rightarrow 1$); otherwise, following a probability $1-f$ to maintain intact. In other words, a memory array --- an array of memory cells --- is randomized by independently applying the RR technique to each bit position through adjusting supply voltages and cell sizing. In so doing, the original data represented and stored in binary format in the memory is sanitized with LDP noises. There are a few challenges in developing \texttt{SRAM\_DP}: (i) how to ensure that the perturbation on binary values preserve LDP on their original values (e.g., decimal, categorical values); (ii) how could the curator revert useful statistics from a large number of sanitized data with high utility; (iii) how to reduce system overhead such as latency in control signaling. This work addresses these challenges; and in summary, we push the frontier of the state-of-the-art as follows.
\begin{enumerate}
  \item This is the first work that simultaneously realizes LDP in hardware (thereby achieving privacy-by-design) and improves power efficiency and system responsiveness, which promises a ``win-win'' outcome.
  \item The data is transformed from its original type to the binary domain for LDP randomization. Theoretical analysis are provided for its rigorous bounds in privacy, utility, and side-channel leakage. 
  \item We design and implement a combination of hardware techniques --- voltage down-scaling and custom memory design --- to introduce controlled noise to the binary data. The developed hardware complies to LDP, preserves utility by protecting significant bits from failure, and reduces power consumption and latency.
\end{enumerate}

To the best of the authors’ knowledge, the proposed hardware-based LDP realization has made the first and interdisciplinary attempt to exploit low-power memory design for LDP. Also, other than the presented SRAM-based scheme presented in this paper, the proposed technique can be implemented using different memory technologies, such as DRAM and nonvolatile memories \cite{fu2020memristor}.

The rest of the paper is organized as follows. Section \ref{pre} provides preliminaries on LDP and SRAM failure characteristics. The details of \texttt{SRAM\_DP} are described in Section \ref{main}. We present the theoretical study on privacy and utility in Section \ref{analysis}. The statistical recovery algorithms are then presented in Section \ref{rec}. Section \ref{eval} demonstrates and discusses simulation and experiment results, which is followed by a discussion on \texttt{SRAM\_DP} reliability issues in Section \ref{reliability}. Finally, Section \ref{relate} surveys related works and Section \ref{con} concludes the paper.

\section{Preliminaries}\label{pre}
\subsection{Local Differential Privacy}
LDP enriches the conventional differential privacy framework by ensuring the \emph{indistinguishability} of two singleton databases (i.e., two arbitrary records) via a local randomization mechanism. The definition of LDP and its enabling technique are detailed as follows.

\begin{definition}[Local Differential Privacy \cite{erlingsson2014rappor}]
LDP allows data contributor to locally perturb its own data using a randomization technique $\mathcal{M}$ that satisfies $\varepsilon$-LDP, before sharing data to the data curator. Specifically, a randomized mechanism $\mathcal{M}$ satisfies $\varepsilon$-LDP, if and only if for two data records $v_{1}$ and $v_{2}$ and for all output $S \subseteq Range(\mathcal{M})$, the following inequality holds:
\begin{equation*}
\Pr [ {{\mathcal{M}} (v_{1})} \in S]  \le {e^\varepsilon } \Pr [ {{\mathcal{M}}( v_{2} )  \in S]}
\end{equation*}
\end{definition}

To achieve LDP, it has been proven that the classic \emph{random response} technique that is widely used in surveys of people's ``yes'' or ``no'' answers for sensitive questions can be adapted to achieve LDP. Its definition is as follows.
\begin{definition}[Randomized Response \cite{warner1965randomized}]
For a private binary value $x \in \{0, 1\}$, a randomized response (RR) scheme follows a $2\times2$ design matrix to perturb $x$, that is $p_{sv} = p[ y = s | x = v ]$ ($s, v \in \{0, 1\}$) as the probability of the output being $s$ when the input is $v$. The RR that satisfies $\varepsilon$-LDP follows that $p_{00} = p_{11} = \frac{e^{\varepsilon}}{1+e^{\varepsilon}}$ and $p_{01} = p_{10} = \frac{1}{1+e^{\varepsilon}}$.
\end{definition}
To handle more general data types, Google developed RAPPOR \cite{erlingsson2014rappor} which applies RR to the Bloom filter encoded data.

\subsection{SRAM Failure Characteristics}\label{fail_cha}
Memories are ubiquitous in today’s electronic systems. Among different memories, SRAM is mainly used as the cache and internal registers of a CPU thanks to its fast read/write speed. SRAM is volatile; its stored data is lost when the supply voltage is removed. This unique property has invited  researchers to investigate the voltage down-scaling technique, a.k.a., low-power memory design, to balance power consumption and data integrity \cite{Xu2020Model}. 

\begin{figure}
\centering
  \includegraphics[width=\linewidth]{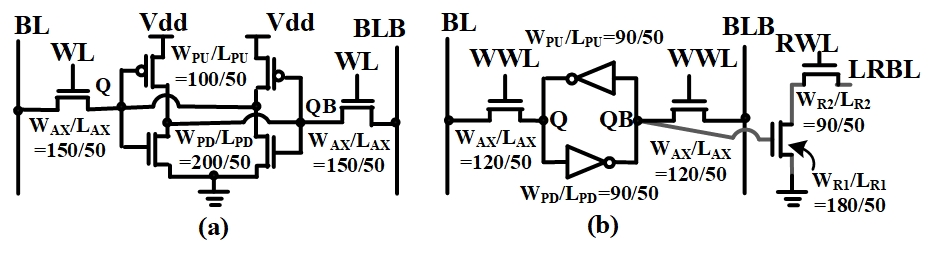}
  \caption{45nm SRAM cell schematics with the smallest silicon area: (a) 6T (C61) and (b) 8T (C81).}
  \label{fig:bitcell_intro}
\end{figure}

In addition to supply voltage, the SRAM failure characteristics also depend on its cell structure and transistor sizes. Specifically, among various SRAM designs, 6T and 8T are the two most widely-applied SRAM cells. Figure~\ref{fig:bitcell_intro} shows 6T and 8T cells with the smallest silicon area in a 45 nm CMOS technology. 6T can enable compact memory design for its small silicon area while 8T can effectively reduce memory failures thanks to the decoupled read and write paths \cite{Gong2012Cell}. At low voltages, SRAM failures are mainly caused by process variations, in particular threshold voltage variations ($\sigma_{Vth}$), which can be expressed as  $\sigma_{Vth} = \frac{A_{VT}}{\sqrt{WL}}$ \cite{croon2004physical}. $A_{VT}$ is a technology dependent constant, and \emph W, \emph L respectively represent the width and length of the transistors. As the \emph W and \emph L of transistors increase, the threshold voltage variations are reduced which makes the SRAM more reliable in retaining its data at low voltages. In other words, upsizing SRAM cells avoids data losses under low voltages but comes with a tradeoff in silicon area. 

\begin{figure}
\centering
  \includegraphics[width=\linewidth]{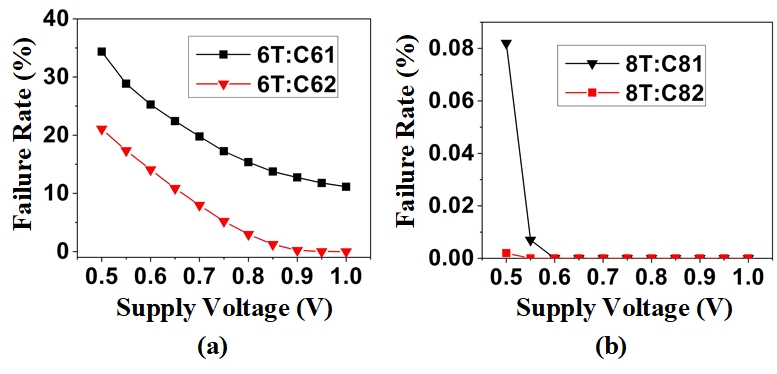}
  \caption{Failure characteristics of 45 nm 6T and 8T cells.}
  \label{fig:failure_rate}
\end{figure}

In our experiments, we show in Figure~\ref{fig:failure_rate} the failure characteristics of two 6T and two 8T cells with different transistor sizes. The detailed silicon area data of four cells is listed in Table I. As compared to the smallest 6T design, i.e., C61, the area overhead of the upsized 6T (C62), the smallest 8T (C81), and the upsized 8T (C82) is 15.4\%, 9.6\%, and 14.3\%, respectively. In our analysis, 100,000 HSPICE Monte-Carlo simulations are performed in the worst process corner to obtain the failure rates of cells, i.e. “fs” (fast NMOS and slow PMOS) for 6T and “sf” (slow NMOS and fast PMOS) for 8T. As shown in Figure~\ref{fig:failure_rate}, the failure rate of a memory is monotone decreasing with respect to (w.r.t.) voltage and area. As also observed, with similar silicon area, 8T cell (C82) has a significant lower failure rate than upsized 6T cells (C62). 

\begin{table}[h!]
\centering
\caption{Silicon areas of 45 nm memory cells.}
\begin{tabularx}{\columnwidth}{X X X X X} 
 \hline
 \small Different Cells & \small Height ($\mu$m) & \small Width ($\mu$m) & \small Area ($\mu$$m^2$) & \small Area Ratio \\ 
 \hline\hline
 6T:C61 & 0.45 & 1.523 & 0.685 & 1.000\\ 
 6T:C62 & 0.45 & 1.758 & 0.791 & 1.154\\
 8T:C81 & 0.45 & 1.663 & 0.751 & 1.096 \\
 8T:C82 & 0.45 & 1.740 & 0.783 & 1.143 \\
 \hline
\end{tabularx}
\label{table:1}
\end{table}

\section{Privacy by Design in SRAM Memory}\label{main}
In this paper, we propose to utilize the SRAM failures at low voltages to inject LDP noises to the stored data, thereby achieving privacy by design while reducing system power consumption. This idea creates a ``win-win'' effect between privacy preservation and power efficiency --- reminiscent of the old-saying: ``kill two birds with one stone''. However, existing SRAM chips cannot accomplish the goal for two reasons. First, a failed memory cell cannot generate a random binary value. In theory, for any bit  $x \in \{0,1\}$ that is stored in a failed cell, it becomes ambiguous to determine implying that 0 and 1 are the possible readouts. However, in practice, for a specific SRAM chip after fabrication, its failed bits will always be readout as 0 or 1. For example, at low voltages, the failed bits of Cypress's commercial memory chip (CY62146GN) consistently generate 1s \cite{Das2021ECC}. This property, coined as \emph{``fixed output upon cell failure''} is unfortunately unacceptable for achieving LDP. Second, SRAM cell failure exhibits a \emph{``fault inclusion''} property; the cells that fail at voltage $V_{1}$ will certainly fail at a lower voltage $V_{2}$ where $V_{2} < V_{1}$ \cite{GOTTSCHO2015fault}. This deterministic pattern will reveal side-information to adversaries who may conduct cell-failure profiling experiments and eliminate unlikely bit values. In this paper, we develop a new memory architecture to address the above two challenges thus accomplishing the vision of \texttt{SRAM\_DP}.

\subsection{Principle of \texttt{SRAM\_DP}}\label{principle}
Given a user's value which could be in any format such as numbers, characters and categories, it is converted to binary bit strings $X = \{x_{1},...,x_{n} \}$ and then stored at the user device's SRAM cells $C = \{c_{1},...,c_{n} \}$. Whenever the user needs to send her data to a curator, $X$ is read out from the memory as bit strings $O = \{o_{1},...,o_{n} \}$ with LDP noise. The overview of the proposed \texttt{SRAM\_DP} mechanism is shown below, which consists of four steps. 

\begin{enumerate}
  \item \textbf{Bit Shift}. User's memory system configures a finite set of permutation vectors as $\Pi = \{\pi_{1},..., \pi_{m} \}$ offline. Before writing a value $X$ to memory for storage, a permutation vector $\pi_{i}$ is randomly selected from $\Pi$ to re-organize the bit-string $X$ which is then stored in the memory. The selected permutation vector is also stored in memory for data reserve-shift.
  \item \textbf{Memory Storage}. The noise injection process is done by enabling low voltages for the custom memory. As discussed in Section \ref{fail_cha}, user's low-voltage memory system generates cell failures across the memory. At a specific voltage, the failure positions can be determined by its built-in self test (BIST) process. 
  \item \textbf{Noise Injection}. Upon a memory read access, depending on the position of failed cells, the readout bits on non-failed cells are kept the same while the readout bits on failed cells are replaced by 0 or 1 randomly.
  \item \textbf{Reverse Bit Shift}. User's memory system re-applies the permutation vector $\pi_{i}$ used in Step 1 to revert the generated bit string into $O$ as the output.
\end{enumerate}

\begin{figure}
\centering
  \includegraphics[width=0.88\linewidth]{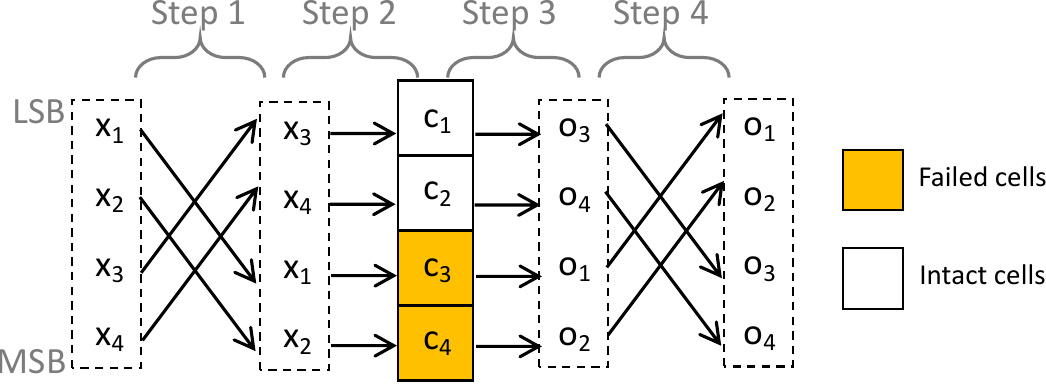}
  \caption{A toy example showing the procedures of \texttt{SRAM\_DP}.}
  \label{toy}
\end{figure}

To put the above steps into perspective, we present a toy example in Figure~\ref{toy} that shows how a 4-bit data $X$ is perturbed. The objective of each step is narrated as follows.

The \textbf{Bit Shift} in Step 1 serves for two purposes. First, permutating input $X$ will address the \emph{``fault inclusion''} problem and effectively defend against side-channel attacks. Second, by shifting bit positions of input $X$, we achieve fine-grained control over the failure rate of each bit position for high utility preservation. Specifically, the most significant bit (MSB) (resp. the least significant bit (LSB)) could be shifted to the cell with least (resp. highest) failure rate, as shown in Figure~\ref{toy}. 

The \textbf{Memory Storage} in Step 2 is to let memory cells manipulate the bits stored therein. Specifically, by online adjusting the voltage $V$ and offline designing the set of permutation vectors $\Pi$, we have control over a key parameter $f$ in LDP; that is, a bit $x_{i} \in X$ will maintain its original value with probability $1-f_i$ (i.e., by being stored in an intact memory cell) while staying in an ambiguous state with probability $f_i$ (i.e., by being stored in a failed memory cell). Mathematically, the output of this step is
\[ o_{i} = \begin{cases}
      0 & \text{with probability} \, \frac{1}{2} f_i \\
      1 & \text{with probability} \, \frac{1}{2} f_i\\
      x_{i} & \text{with probability} \, 1 - f_i
   \end{cases}
\]

The \textbf{Noise Injection} in Step 3 is to address the \emph{``fixed output upon cell failure''} issue. During the readout process, we inject random noises (0 or 1) to the failed cell positions. 

The \textbf{Reverse Bit Shift} in Step 4 is to recover the original bit positions of $X$. 

\subsection{Hardware Architecture of \texttt{SRAM\_DP}}
\begin{figure*}
  \centering
  \includegraphics[width=\linewidth]{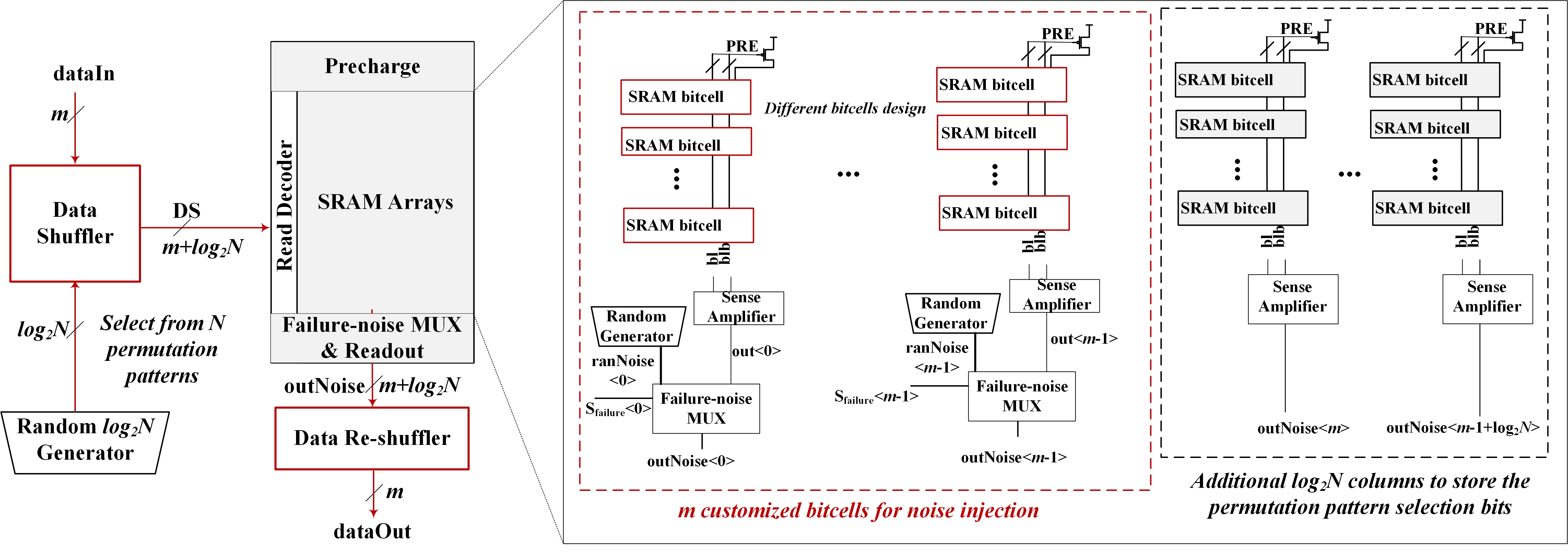}
  \caption{Hardware Architecture of the proposed \texttt{SRAM\_DP}.}
  \label{fig:memory_overview}
\end{figure*}

To support the above procedures in our proposed \texttt{SRAM\_DP}, we re-design the conventional SRAM memory architecture into a novel one which is shown in Figure~\ref{fig:memory_overview}.

To protect privacy, the failed bits in SRAM would be used to add random binary bits as noise. As shown in Figure~\ref{fig:memory_overview}, a data shuffler is added to implement the \textbf{Bit Shift} in Step 1, which is designed based on the configured permutation vectors. The random bit generation logic for the \textbf{Noise Injection} in Step 3 can be efficiently implemented by connecting multiplexers (MUX) to sense amplifier (SA) readout values of conventional SRAM. Each SA readout bit is connected to a MUX which is controlled by the received memory failure positions. If a memory failure is indicated, a random binary bit is enabled by selecting output of random bit generator. The failure position information will be used as the select signal of the MUX to control which bit to be the output. Similar to other existing fault position aware mitigation techniques \cite{edstrom2017data}, the proposed \texttt{SRAM\_DP} receives pre-determined locations of failed bits, which is usually executed either during post fabrication testing or during power-on self-test (POST). Finally, a data reshuffler is added to realize the \textbf{Reverse Bit Shift} in Step 4, based on the permutation pattern selection bits stored in the memory. The detailed hardware implementations and evaluation results of \texttt{SRAM\_DP} will be discussed in Appendix B and Section \ref{eval}, respectively. 

\section{Privacy and Utility Analysis}\label{analysis}
\subsection{Compliance to LDP Notion}
First of all, we attempt to draw a connection between the parameter $f$ and the memory behaviors of \texttt{SRAM\_DP}. It is clear that whether a bit $x_{i} \in X$ can maintain its original value, i.e., $o_{i} = x_{i}$, is determined by (i) the bit shift pattern and (ii) memory cell failure rate. For the former factor, suppose that a randomly selected permutation vector $\pi$ could shift a bit $x_{i}$ to any memory cell $c_{k}$ following the probability vector $\bm{P}_{\pi,x_{i}} = \{p_{x_{i} \rightarrow c_{1}},...,p_{x_{i} \rightarrow c_{n}}\}$ where $\sum_{k=1}^{n} p_{x_{j} \rightarrow c_{k}} = 1$ and $x_{i} \rightarrow c_{k}$ denotes the event that bit $x_{i}$ is shifted to (or stored at) the memory cell $c_{k}$. For the latter factor, when a specific voltage V is applied, a memory will exhibit the failure characteristics similar to Figure~\ref{fig:failure_rate}. Suppose that a memory cell $c_k$ exhibits a failure probability $p_{V,c_k}$ at voltage V. By taking these two factors into account, the probability that a bit retains its original value because of being stored in an intact cell is
\begin{equation*}
P(o_{i} = x_{i}) = \sum_{k=1}^{n} p_{x_{i} \rightarrow c_{k}} (1-p_{V,c_k}) = 1 - \sum_{k=1}^{n} p_{x_{i} \rightarrow c_{k}} p_{V,c_k}.
\end{equation*}
By relating the above equation to $f$, we can conclude that $f_i = \sum_{k=1}^{n} p_{x_{i} \rightarrow c_{k}} p_{V,c_k},$ $\forall x_{i} \in X$.

Next, we present the relationship between $f$ and $\epsilon_{\infty}$ in the following theorem.
\begin{theorem}\label{DP}
The \texttt{SRAM\_DP} mechanism satisfies $\epsilon_{\infty}$-differential privacy for any input bit strings $X$, where $\epsilon_{\infty}$ = $ \sum_{i=1}^{n} \text{ln}(\frac{1 - \frac{1}{2}f_{i}}{\frac{1}{2}f_{i}})$.
\end{theorem}
\begin{proof}
Without loss of generality, assume $X_{1} = \{x_{1} = 0,...,x_{n} = 0\}$ and $X_{2} = \{x_{1} = 1,...,x_{n} = 1\}$ meaning that $X_{1}$ and $X_{2}$ are complimentary and the $l_{1}$ norm is maximized. Besides, we know that for any bit $x_{i} \in X$, \texttt{SRAM\_DP} generates the readout bit $o_{i} \in O$ with probability
\begin{align}\label{RR_prob}
\begin{split}
&P(o_{i} = 1 \, | \, x_{i} = 1) = \frac{1}{2}f_{i} + 1 - f_{i} = 1 - \frac{1}{2}f_{i}, \\
&P(o_{i} = 1 \, | \, x_{i} = 0) = \frac{1}{2}f_{i}.
\end{split}
\end{align}
Similarly, $P(o_{i} = 0 \, | \, x_{i} = 0) = 1 - \frac{1}{2}f_{i}$ and $P(o_{i} = 0 \, | \, x_{i} = 1) = \frac{1}{2}f_{i}$. Then, for any output $O^{*} = \{o_{1},...,o_{n}\}$,
\begin{align*}
P(O = O^{*} \, | \, X = X_{1}) =& (\frac{1}{2}f_{1})^{o_{1}} \cdot (1 - \frac{1}{2}f_{1})^{1 - o_{1}} \times ... \\
& \times (\frac{1}{2}f_{n})^{o_{n}} \cdot (1 - \frac{1}{2}f_{n})^{1 - o_{n}}
\end{align*}
\begin{align*}
P(O = O^{*} \, | \, X = X_{2}) =& (1 - \frac{1}{2}f_{1})^{o_{1}} \cdot (\frac{1}{2}f_{1})^{1 - o_{1}} \times ... \\
& \times (1 - \frac{1}{2}f_{n})^{o_{n}} \cdot (\frac{1}{2}f_{n})^{1 - o_{n}}
\end{align*}

Consider that the \texttt{SRAM\_DP} design abstractly provides a randomization function $\mathcal{M}$ where output $\mathcal{M}(X) \in S$. For the LDP condition to hold, we need to ensure the ratio of probabilities of getting the same output for any two arbitrarily ``adjacent'' inputs $X_1$ and $X_2$ to be bounded by exp($\epsilon_{\infty}$). 
Mathematically, we have
\begin{align*}
\begin{split}
\frac{P(\mathcal{M}(X_1) \in S)}{P(\mathcal{M}(X_2) \in S)} =& \frac{P(O = \mathcal{M}(X_1) \, | \, X = X_{1})}{P(O = \mathcal{M}(X_2) \, | \, X = X_{2})} \\
=& \frac{ \sum_{O^{*} \in S} P(O = O^{*} \, | \, X = X_{1})}{ \sum_{O^{*} \in S} P(O = O^{*} \, | \, X = X_{2})} \\
\leq & \max_{O^{*} \in S} \frac{P(O = O^{*} \, | \, X = X_{1})}{P(O = O^{*} \, | \, X = X_{2})} \\
=& \, (\frac{1}{2}f_{1})^{2o_{1} - 1} \cdot (1 - \frac{1}{2}f_{1})^{1 - 2o_{1}} \times ... \\
& \times (\frac{1}{2}f_{n})^{2o_{n} - 1} \cdot (1 - \frac{1}{2}f_{n})^{1 - 2o_{n}} \\
=& \prod_{i=1}^{n} (\frac{\frac{1}{2}f_{i}}{1 - \frac{1}{2}f_{i}})^{2o_{i} - 1} \\
\leq & \prod_{i=1}^{n} (\frac{1 - \frac{1}{2}f_{i}}{\frac{1}{2}f_{i}}) = exp(\epsilon_{\infty})
\end{split}
\end{align*}
where the second inequality holds when every $o_{i} = 0$. The reason is as follows. First of all, it is clear that $\frac{\frac{1}{2}f_{i}}{1 - \frac{1}{2}f_{i}}$ takes values from $[0,1]$ because $f_{i} \in [0,1]$. Then, provided that $o_{i}$ is binary (0 or 1), the exponent ${2o_{i} - 1}$ is either -1 or 1, respectively. Therefore, the upper bound of the above equation occurs when $o_{i} = 0$, $\forall i$. 
\end{proof}

\subsection{Side-channel Analysis}
In addition to proving that \texttt{SRAM\_DP} guarantees LDP, it is equally vital to examine if \texttt{SRAM\_DP} leaks any side-channel information that may jeopardize data privacy. Amongst the designed components of \texttt{SRAM\_DP}, the power control unit --- a firmware executed by CPU --- for voltage scaling is vulnerable to (remote) power profiling attacks \cite{wang2022hertzbleed,dipta2022df}. For instance, Hertzbleed threat allows attackers to steal full cryptographic keys by observing variations in CPU frequency enabled by dynamic voltage scaling \cite{wang2022hertzbleed}. Nonetheless, in our \texttt{SRAM\_DP} design, the voltage profile is only correlated with $\epsilon$ as shown in the above proof and Table~\ref{tab:fail:dp}. Knowing $\epsilon$ gives no benefits to attackers --- $\epsilon$ by default is a public information, not to mention that extracting $\epsilon$ is almost impossible because attackers have no clues of victim devices' SRAM specifications such as its transistor sizing, as illustrated earlier in Figure~\ref{fig:failure_rate}. 

Even for the most adversarial case where attackers claim root access to CPU and gain full knowledge of SRAM specifications, LDP noises cannot be stolen or in any way inferred because they are generated and utilized on-the-fly during the SRAM read/write process (achieved by the sense amplifier and noise generator in Figure~\ref{fig:memory_overview}) and they are never stored. Moreover, the sensitive data are naturally perturbed when they are written into memory. Therefore, any attempt in reading data from SRAM will not be able to reverse their true values.

\subsection{Utility Analysis}
\begin{theorem}\label{DP2}
The expected $l_1$ utility loss of \texttt{SRAM\_DP}, denoted as $\mathbb{E}(|O-X|)$, has a ${\cal O}(c^n)$ upper bound where $c$ is a constant relevant to $\{f_{1},...,f_{n}\}$.
\end{theorem}
\begin{proof}
Denote the change of bit value as $\Delta a_{i}=o_{i}-x_{i} \in \{-1,0,1\}$. Without any prior assumption on $x_{i}$, i.e., it can take value 0 or 1 with equal probability, the probability of $\Delta a_{i}$ is given as 
\begin{equation} \label{pmf}
P(\Delta a_{i}) = \left\{ \begin{array}{l}
\frac{1}{4}f_{i}, \quad \quad \quad \Delta a_{i} = -1 \\
1-\frac{1}{2}f_{i}, \quad \, \, \, \Delta a_{i} = 0  \\
\frac{1}{4}f_{i}, \quad \quad \quad \Delta a_{i} = 1
\end{array} \right.
\end{equation}
according to Eq.(\ref{RR_prob}). Upon mapping randomization in binary domain back to its decimal format, we can obtain the change of data’s original value $\Delta A \mathop = \limits^{\text{def}} O-X = \Delta a_{n-1} 2^{n-1}+\Delta a_{n-2} 2^{n-2}+\cdots+\Delta a_{0} 2^{0}$. $\Delta A$ is a random variable whose value is an integer from $\{-2^{n}+1,...,0,...,2^{n}-1\}$. Then, to derive the expected value of $|O-X|$ (or $|\Delta A|$) which is $l_1$ loss of \texttt{SRAM\_DP}, we have the following Lemma for the probability mass function (PMF) of $\Delta A$.

\begin{lemma}\label{DP2_lemma}
The PMF of $\Delta A$ is zero-mean and symmetric w.r.t. $\Delta A = 0$. The proof is shown in the Appendix A.
\end{lemma}

With Lemma \ref{DP2_lemma}, we can derive $\mathbb{E}(\Delta A)$ as follows:
\begin{align*}
\mathbb{E}(|O-X|) = \sum_{|O-X|=0}^{2^{n}-1} |O-X| P(|O-X|) 
\end{align*}
\begin{align*}
\begin{split}
&\mathop  = \limits^{(1)} 2 \sum_{\Delta A=0}^{2^{n}-1} \Delta A  P(\Delta A) \\
&\mathop  = \limits^{(2)} 2 \sum_{\Delta A=0}^{2^{n}-1} \Delta A   \sum_{K \in {\cal K}_{\Delta A}} \left[\prod_{i \in I} \left(1-\frac{1}{2} f_{i}\right) \prod_{i \in I^\prime} \left(\frac{1}{4} f_{i}\right) \right] \\
&\mathop  \leq \limits^{(3)} 2 \sum_{\Delta A=0}^{2^{n}-1} \Delta A  \sum_{k=0}^{n} \left[\prod_{i = 0}^{k} \left(1-\frac{1}{2} f_{i}\right) \prod_{j = k+1}^{n} \left(\frac{1}{4} f_{i}\right) \right] \\
&\mathop = \limits^{(4)} 2^{2n-1} \sum_{k=0}^{n} \left[\prod_{i = 0}^{k} \left(1-\frac{1}{2} f_{i}\right) \prod_{j = k+1}^{n} \left(\frac{1}{4} f_{j}\right) \right]
\end{split}
\end{align*}
We have equality (1) because of Lemma \ref{DP2_lemma}. The rationale of equality (2) is that any value of $\Delta A$ is due to a combination of $\left\{\Delta a_{n-1}, \ldots, \Delta a_{0}\right\}$. For example, for a three-bit ($n$=3) decimal number, there are two combinatorics that can result in $\Delta A = 5$, which are ${\cal K}_{\Delta A = 5} = \left\{ \{1,1,-1 \}, \{1,0,1 \} \right\}$. Here, ${\cal K}_{\Delta A}$ represents the set of combinatorics that lead to $\Delta A$. Any combinatoric $K$ in ${\cal K}_{\Delta A}$ is a specific realization of $\Delta A$. For any $K$, it contains $n$ corresponding values (e.g., $K(j)$ is the $j^\text{th}$ element) in $\left\{\Delta a_{n-1}, \ldots, \Delta a_{0}\right\}$. Then, the probability of $\Delta A = a$, $\forall a \in \{-2^{n}+1,...,0,...,2^{n}-1\}$, is simply the addition of probabilities for having every combinatorics in ${\cal K}_{\Delta A = a}$. More specifically, the probability of any combinatoric $K$ in ${\cal K}_{\Delta A = a}$ is the product of probabilities in Eq.(\ref{pmf}) for all $\left\{\Delta a_{n-1}, \ldots, \Delta a_{0}\right\}$. For notation simplicity, denote $I = \{j|K(j) = 0, 1 \leq j \leq n\}$ and $I^\prime = \{j|K(j) \neq 0, 1 \leq j \leq n\}$. For instance, in the above three-bit example, $P(\Delta A = 5)$ = $\left(\frac{1}{4} f_1\right)$ $\left(\frac{1}{4} f_2\right)$ $\left(\frac{1}{4} f_3\right)$ + $\left(\frac{1}{4}f_1\right)$ $\left(1-\frac{1}{2} f_2\right)$ $\left(\frac{1}{4} f_3\right)$.

Note that deriving ${\cal K}_{\Delta A}$ for all possible $\Delta A$ requires exhaustive search which is significantly time-consuming as the search space increases exponentially w.r.t. $n$. Nonetheless, it is intuitive to assert that for any specific value of $\Delta A$, as long as the number of zero-valued elements in $\left\{\Delta a_{n-1}, \ldots, \Delta a_{0}\right\}$ are fixed, the remaining non-zero elements will be deterministic. In light of it, we have inequality (3) by relaxing the carnality of ${\cal K}_{\Delta A}$ to $n$ --- the maximum number of zero-valued elements. Moreover, without loss of generality, we can neglect the order of cell failure positions $\{f_{1},...,f_{n}\}$ and simply consider the number of failed cells that lead to $\Delta a_* = 0$ (with probability $1-\frac{1}{2} f_{*}$). Then, we can easily conclude with the result by calculating the sum of arithmetic series in equality (4).

Obviously, the expected $l_1$ utility loss of \texttt{SRAM\_DP} has an ${\cal O}(c^n)$ upper bound, in which $c$ is a constant relevant to $\{f_{1},...,f_{n}\}$. In fact, this claim will become more intuitive at a special case of homogeneous failure rate $f_{1}=...=f_{n} = f$, for which the above derivation can be further simplified into
\begin{equation*}
\mathbb{E}(|O-X|) \leq \left(4^{n}-2^{n}\right) \cdot \frac{\left(1-\frac{1}{2} f\right)^{n+1}-\left(\frac{1}{4} f\right)^{n+1}}{1-\frac{3}{4} f}.
\end{equation*}

All the above analytical results of Lemma \ref{DP2_lemma} will later be validated numerically in Figure~\ref{fig:utility}.
\end{proof}

\section{Statistics Recovery Algorithms}\label{rec}
In this work, we assume each user's data is independently and identically distributed (i.i.d.) with a specific distribution $\mathcal{P}$. After applying \texttt{SRAM\_DP}, a data curator needs to recover certain statistics of $\mathcal{P}$. In this section, we adopt two statistics recovery algorithms, namely the expectation-maximization (EM) algorithm and the constrained linear regression (CLR) algorithm, to achieve the goal. Note that EM and CLR are well-established methods and the purpose of selecting them is to convey the idea that the hardware-perturbed data by \texttt{SRAM\_DP} can be easily analyzed by classical recovery algorithms.

Before running recovery algorithms, the data curator should (i) be aware and apply the same encoding algorithm that is used at users' SRAM system to convert data of arbitrary format into bit strings, (ii) obtain every user's failure rate vector $F = \{f_1,...,f_n\}$ thus the LDP parameter $\epsilon$, and (iii) construct a set of data candidates $\Omega$ that cover all the possible values of users' choices. For instance, for heart beat monitoring, a reasonable heart beat range is between 30bpm to 150bpm so $\Omega = \{X \, | \, 30 \leq X \leq 150, X \in \mathbb{Z^{+}} \} $ and a user's specific input $X \in \Omega$. These requirements could be easily fulfilled by relying on users to piggyback them with the sanitized data or the curator retrieving them from side channels without compromising the differential privacy condition.
\subsection{EM Algorithm}
The essence of EM algorithm is to find the maximum-likelihood estimates (MLEs) through iteration. In the context of this work, the data curator is interested to search through the candidate set $\Omega$ for the most likely $X^{*}$ that contributes to the observed sanitized data $O$. 
Specifically, our EM algorithm consists of the following steps.
\begin{enumerate}
  \item \textbf{Initialization}. Assume the data curator has no additional prior information about users' data, it will initialize the prior probability by setting a uniform distribution as $P(X) = \frac{1}{| \Omega |}$, $\forall X \in \Omega$.
  \vspace{0.05in}
  \item \textbf{Likelihood Calculation}. When the data curator observes a sanitized bit strings $O = \{o_1,...,o_n \}$ from a user, it calculates the likelihood of generating $O$ by any candidate from $\Omega$. Specifically, for the \texttt{SRAM\_DP} mechanism, an output bit $o_i \in O$ maintains its original value with probability $1 - \frac{1}{2}f_{i}$ whereas flips with probability $\frac{1}{2}f_{i}$ as derived in Eq.(\ref{RR_prob}). Then, for a candidate represented in bit string $X = \{x_1,..., x_n\}$, each of its bit has the likelihood of $P(o_i \, | \, x_i) = (\frac{1}{2}f_{i})^{\beta_i} \cdot (1 - \frac{1}{2}f_{i})^{1 - \beta_i}$ of generating $o_i$ for $\forall o_i \in O$, where $\beta_{i}$ = XOR$(o_i, x_i)$ and XOR$()$ is the boolean exclusive OR operator. Then, for any user's submitted data, we have
  \vspace{-0.05in}
      \begin{equation}\label{cond_prob}
      P(O \, | \, X) = \prod_{i=1}^{n} (\frac{1}{2}f_{i})^{\beta_i} \cdot (1 - \frac{1}{2}f_{i})^{1 - \beta_i}, \quad \forall X \in \Omega.
      \end{equation}

  \item \textbf{Posterior Calculation}. Given all the conditional distributions of one particular observation $O$ as above, the corresponding posterior probability can be calculated according to the Bayesian theorem, which is given below.
  \vspace{-0.15in}
      \begin{equation}\label{post_prob}
      P(X \, | \, O) = \frac{P(X) \cdot P(O \, | \, X)} {\sum_{X \in \Omega} P(X) \cdot P(O \, | \, X)}, \quad \forall X \in \Omega.
      \end{equation}

      \item \textbf{Update and Iteration}. After calculating the posterior for every user, we update the prior probability for the calculation of next round. Specifically, we take the average of posterior probabilities of all $U$ users in the current iteration and assign it to the prior probability used for the next iteration, mathematically,
      \vspace{-0.1in}
      \begin{equation*}
      P(X) = \frac{1}{U} \sum_{m=1}^{U} P(X \, | \, O_m), \quad \forall X \in \Omega
      \end{equation*}
\end{enumerate}
The aforementioned steps continue till convergence. We can set a stopping criteria as $\max_{X \in \Omega} |P_t(X) - P_{t-1}(X)| \leq \delta$.

\subsection{CLR Algorithm}
Another popular algorithm to recover statistics is the empirical estimation method \cite{murakami2019utility}, which computes an empirical estimate $\hat{\mathcal{P}}$ of $\mathcal{P}$ using the empirical distribution $\hat{\mathcal{Q}}$ of the sanitized data $O$ given the LDP randomization matrix (or conditional probability matrix) $\mathcal{M}$. Mathematically, we have $\hat{\mathcal{P}} \mathcal{M} = \hat{\mathcal{Q}}$, whose closed-form solutions can be easily computed and obtained. As the number of users increases, the empirical distribution $\hat{\mathcal{Q}}$ gets asymptotically close to the true distribution of sanitized data $\mathcal{Q}$, thus $\hat{\mathcal{P}}$ also converges to $\mathcal{P}$ given invariant $\mathcal{M}$. However, when the number of users is small, we cannot ensure the positiveness of $\hat{\mathcal{P}}$. One method is to use Bonferroni correction \cite{wang2017locally} to eliminate estimates below a significance level. Here in this work, we will leverage a constrained linear regression model to obtain $\hat{\mathcal{P}}$. The detailed steps are shown as follows.
\begin{enumerate}
  \item \textbf{Construct $\mathcal{M}$}. \texttt{SRAM\_DP} applies a randomization matrix $\mathcal{M} \in \mathbb{R}_{\geq 0}^{|\Omega| \times |\Omega|}$, which characterizes the conditional probability of mapping an input $X$ to a sanitized output $O$. The calculation of $\mathcal{M}$ follows Eq.(\ref{cond_prob}).
  \vspace{0.05in}
  \item \textbf{Calculate $\hat{\mathcal{P}}$}. We first build the empirical distribution $\hat{\mathcal{Q}}$ by normalizing the frequencies of each record. Then, we solve the following constrained least-square problem:
    \begin{equation}
    \begin{aligned}
    & {\min_{\hat{\mathcal{P}}}}
    & & \frac{1}{2}\|\hat{\mathcal{P}} \mathcal{M} - \hat{\mathcal{Q}} \|_{2}^{2} \\
    & \text{s.t.}
    && \mathbb{E} [X^j] = m_j,  \quad j \geq 1; \\
    & & & \quad 0 \leq \hat{\mathcal{P}}(X) \leq 1, \quad \forall X \in \Omega.
    \end{aligned}
    \end{equation}
    to obtain the optimizer $\hat{\mathcal{P}}$ as the estimation for the input distribution $\mathcal{P}$. Note that $\mathbb{E} [X^j]$ is the $j^{\text{th}}$ moment of the probability distribution $\mathcal{P}$, which is the prior knowledge known to the reconstruction algorithm. 
\end{enumerate}

\section{Performance Evaluation}\label{eval}
In this section, we give a thorough evaluation of our design in both simulations and experiments. We first present several metrics for quantitative evaluation. We consider two specific adversarial side information, one of which strongly pertains to our \texttt{SRAM\_DP} mechanism, to examine the absolute privacy level besides the above $\epsilon_{\infty}$-DP theoretical analysis. We then assess how accurately the EM-based and CLR-based algorithms can recover original statistics. Furthermore, based on the simulation results, we implement a proof-of-concept SRAM to achieve \texttt{SRAM\_DP}. We carry experiments to validate its performance w.r.t. the proposed metrics.

\subsection{Evaluation Metrics}
\textbf{Privacy Meter:} The \texttt{SRAM\_DP} design conforms to the differential privacy notion given the proof in Theorem \ref{DP}. In other words, an adversary is unable to distinguish any datum $X$ from its ``neighbouring'' data. Without loss of generality, we denote that such ``neighbouring'' data collectively form an \emph{indistinguishable set}, whose size is determined by the number of failed cells in SRAM --- denoted as $z$. Obviously, an adversary will have to guess the true datum from a larger \emph{indistinguishable set} when there are more failed cells, leading to a higher uncertainty for the adversary. Since different privacy only poses privacy guarantee within the \emph{indistinguishable set}, the size of the \emph{indistinguishable set} that is $2^{z}$ has a strong implication on the adversarial inference accuracy. In this work, we assume that an adversary follows the maximum likelihood estimation (MLE) to obtain the inferred data $\hat{X}$ from its observation $O$. Then, we follow the privacy meter to measure the \emph{in-accurateness} of adversarial inference as $\text{IA} = \sum_{X} P(X \mid O) |\hat{X}-X|$ which is equivalent to the privacy level.

\vspace{0.05in}
\textbf{Utility Meter:}
Privacy always comes with the cost of utility, and our design is no exception. Intuitively, more failed cells or higher cell failure rates lead to more severe data corruption thus lower utility. Moreover, utility is also affected by the positions of failed cells, in the sense that a higher utility is retained when failed cells occur at less significant bits. To characterize utility level of our design, we use $l_1$ loss (i.e., absolute error) to measure the utility loss as $\text{UL} = \sum_{O} P(O \mid X) |O-X|$.

\subsection{Adversarial Side Information}
We assume two specific adversarial knowledge. The simplest one is to assume an oblivious adversary, which possesses no additional information about the original data and has to guess randomly. That is to say, the prior probability in the MLE inference $P(X) = \frac{1}{2^{z}}$. We denote such adversarial knowledge as \textbf{K1}. Furthermore, we consider a more capable adversary that knows the statistics (e.g., range, mean and variance) of the original data set but is unaware of any specific individual's data. This adversarial knowledge is denoted as \textbf{K2}. Then, an adversary with \textbf{K2} can easily derive a more accurate $P(X)$ by calculating the frequencies of specific elements, for instance, the frequency of data that is in binary form of ``111*****'' (i.e., a 8-bit value with three leading 1s in MSBs followed by five wild cards). Note that both \textbf{K1} and \textbf{K2} are aware of the details of \texttt{SRAM\_DP} such as cell failure rates and $\epsilon$.

\subsection{Input Data}
We generate 1,000 8-bit random binary numbers (i.e., of values 0-255) as input data by sampling a Gaussian distribution with $\mu = 125$ and $\sigma = 20$. For notional convenience, we refer every bit using ``LSB*''. For instance, LSB1 indicates the least significant bit whereas LSB1-3 indicates the least three significant bits. 

\subsection{Simulation Results and Analysis}
\textbf{Simulation Setup:} To simulate the algorithmic procedures of \texttt{SRAM\_DP}, we use MATLAB\_R2022a to implement the key steps in Section \ref{principle}. Specifically, we first sample 1,000 random numbers, convert them into binary values, shuffle their bit positions, and follow an i.i.d., Bernoulli multivariate vector (mimicking the SRAM cell failure probabilities) to decide if any binary values need to be changed into an constant value (e.g., 1s, which reflects the \emph{fixed output upon cell failure} property). Next, we replace the prior changed values with 1s or 0s randomly. In the end, we reverse the binary values back to their original bit positions, which completes the \texttt{SRAM\_DP} randomization life-cycle.

\begin{figure}[!t]
\begin{subfigure}[t]{0.235\textwidth}
  \includegraphics[width=\linewidth]{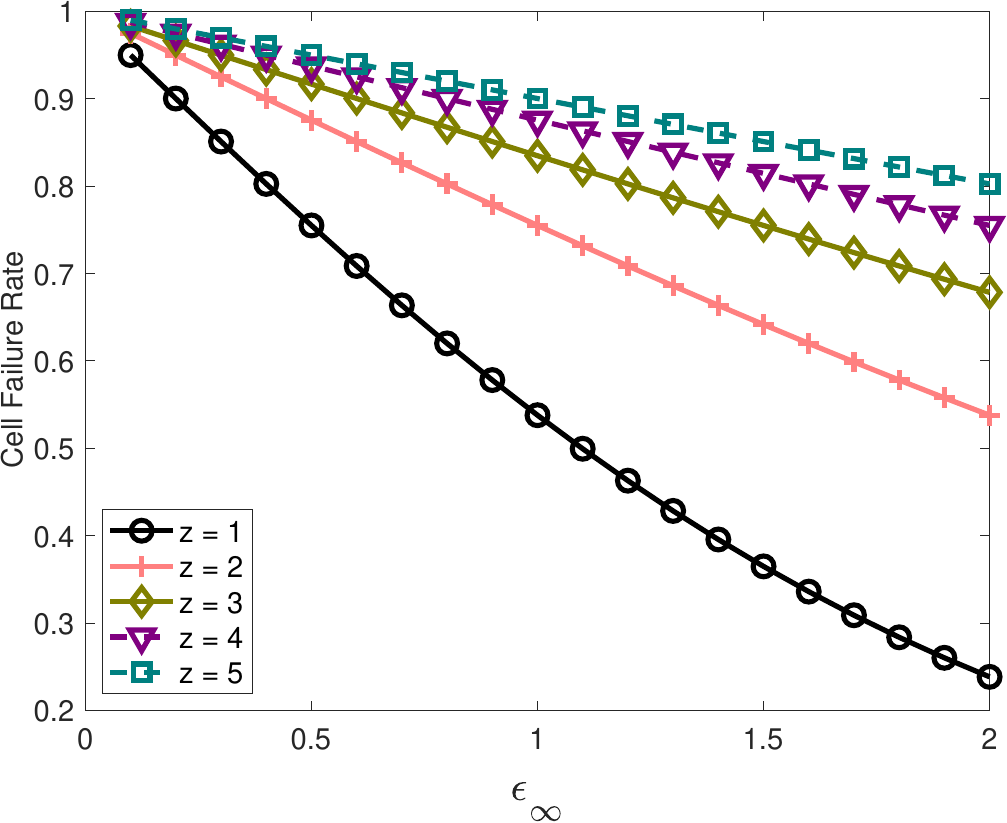}
  \caption{\small Cell failure rates w.r.t. $\epsilon_{\infty}$}
  \label{fig:eps_inf}
\end{subfigure}
\begin{subfigure}[t]{0.235\textwidth}
  \includegraphics[width=\linewidth]{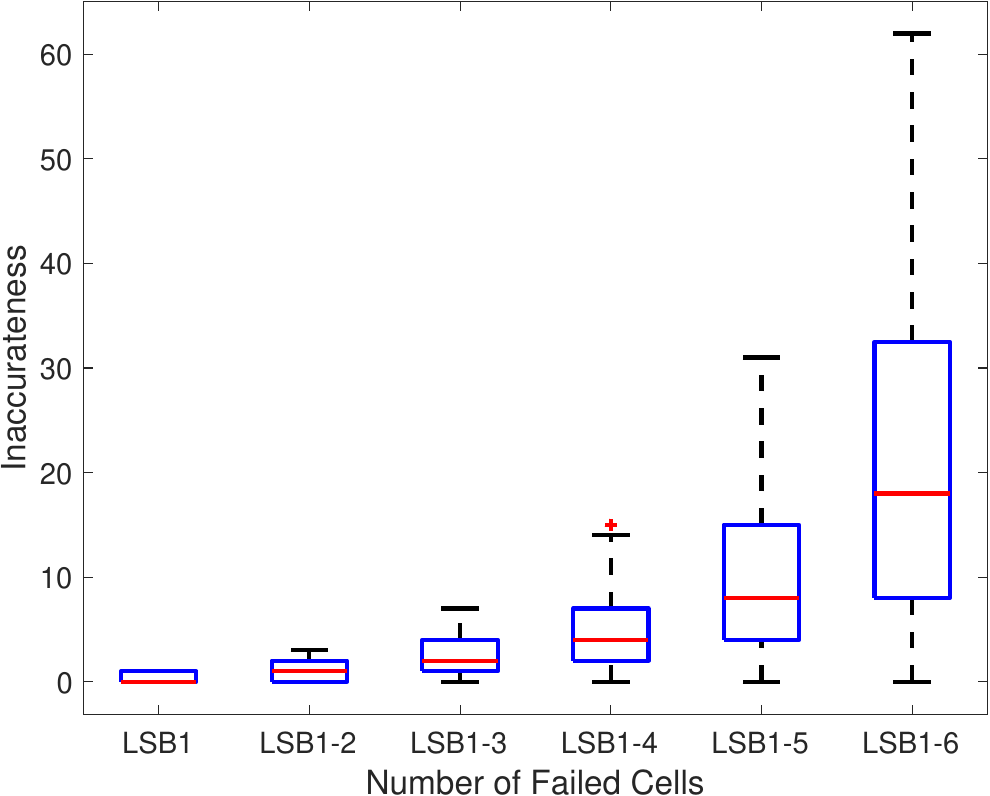}
  \caption{\small IA w.r.t. $z$} \label{ia_num}
\end{subfigure}
   \caption{Privacy analysis.}
\end{figure}

\textbf{Privacy Analysis:} To preserve the $\epsilon_{\infty}$-DP guarantee in Theorem \ref{DP}, we must select appropriate cell failure rates $f$, the number of failed cells $z$, and the positions of failed cells. For the sake of simplicity, we assume homogeneous failure rates among memory cells. Then, we plot how the cell failure rate changes w.r.t. $\epsilon_{\infty}$ as shown in Figure~\ref{fig:eps_inf}.
The result indicates that a stronger privacy (i.e., smaller $\epsilon_{\infty}$) is achieved by increasing cell failure rates and confining failures in a small number of memory cells. For example, a single cell failure with probability 0.5, i.e., $z=1$ with $f=0.5$, yields $\text{ln}3$-DP. Nevertheless, a small \emph{indistinguishable set} for a small $z$ gives rises a higher chance for adversaries to infer the correct value. As shown in Figure~\ref{ia_num}, a larger $z$ offers a higher IA under the adversarial knowledge \textbf{K1}.

\begin{figure}[!t]
\begin{subfigure}[t]{0.235\textwidth}
  \includegraphics[width=\linewidth]{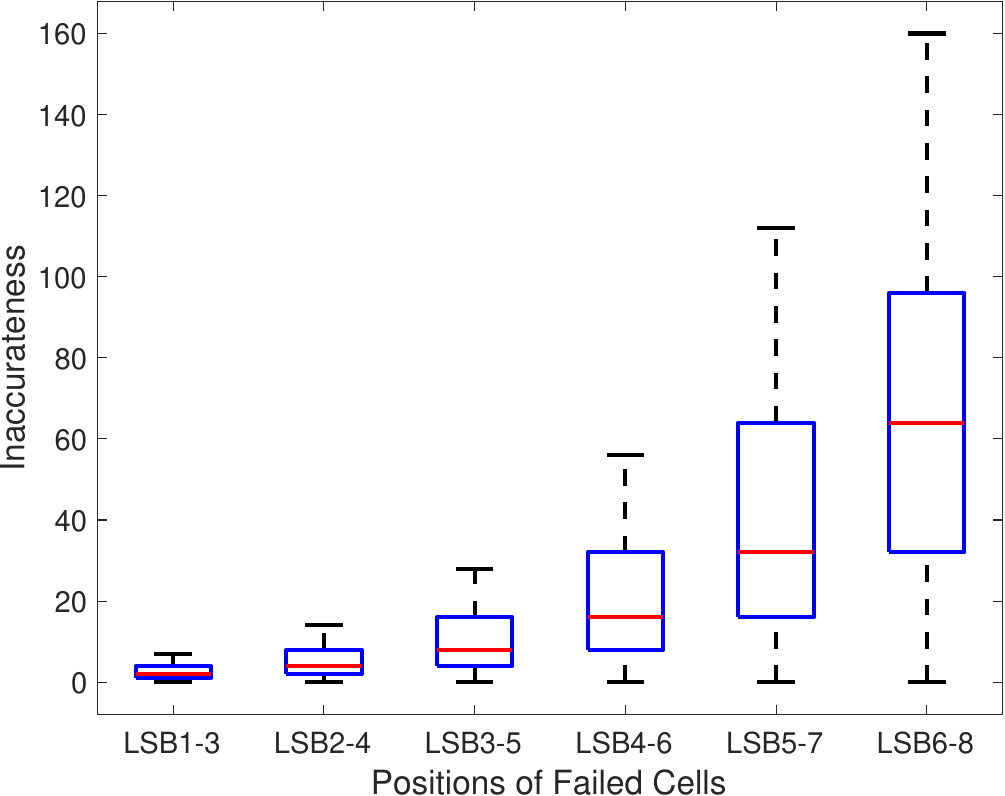}
  \caption{\small IA with prior knowledge \textbf{K1}} \label{ia_wo}
\end{subfigure}
\begin{subfigure}[t]{0.235\textwidth}
  \includegraphics[width=\linewidth]{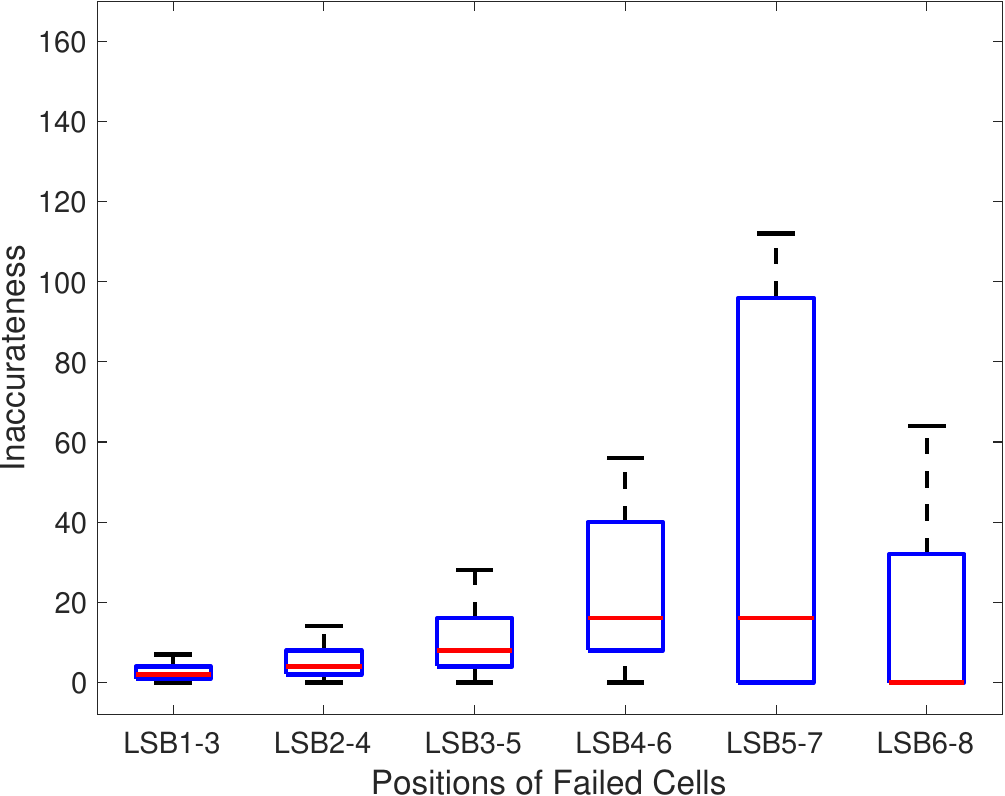}
  \caption{\small IA with prior knowledge \textbf{K2}} \label{ia_w}
\end{subfigure}
   \caption{Privacy level under different prior knowledge.} \label{fig:ia}
\end{figure}

\begin{figure*}[!t]
\centering
\captionsetup{justification=centering}
\begin{subfigure}[t]{0.3\textwidth}
  \includegraphics[width=\linewidth]{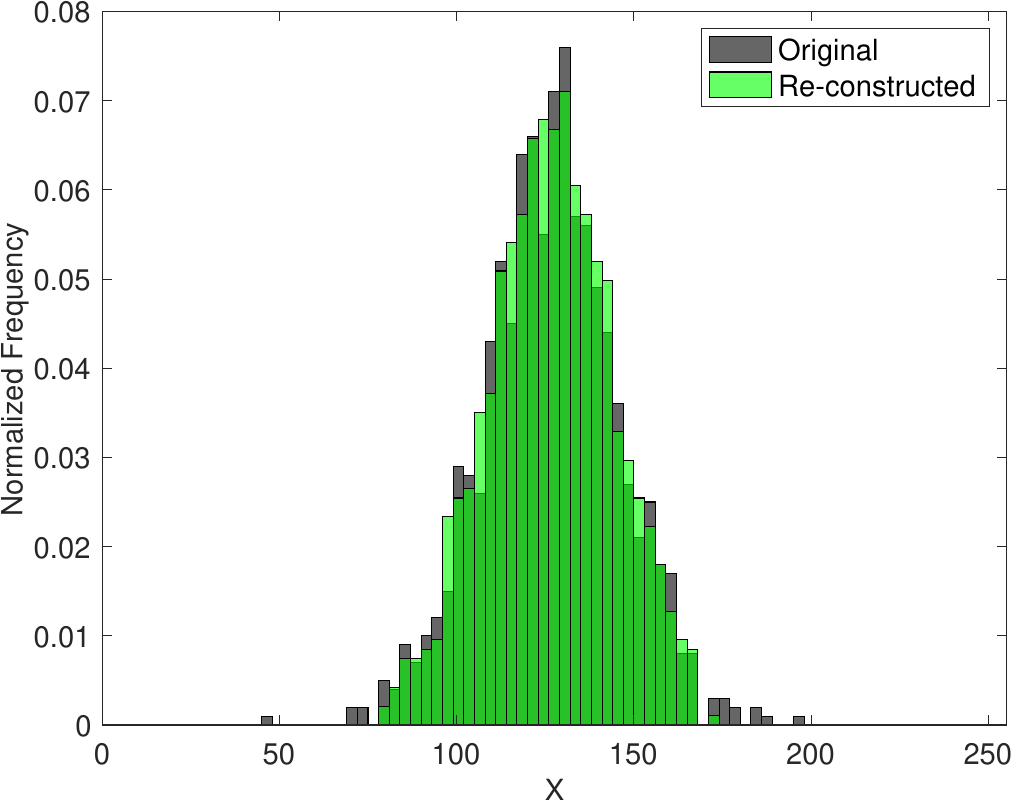}
  \caption{\footnotesize EM-based Re-construction on \textbf{F1}.} \label{em_fp1}
\end{subfigure}\hfill
\begin{subfigure}[t]{0.3\textwidth}
  \includegraphics[width=\linewidth]{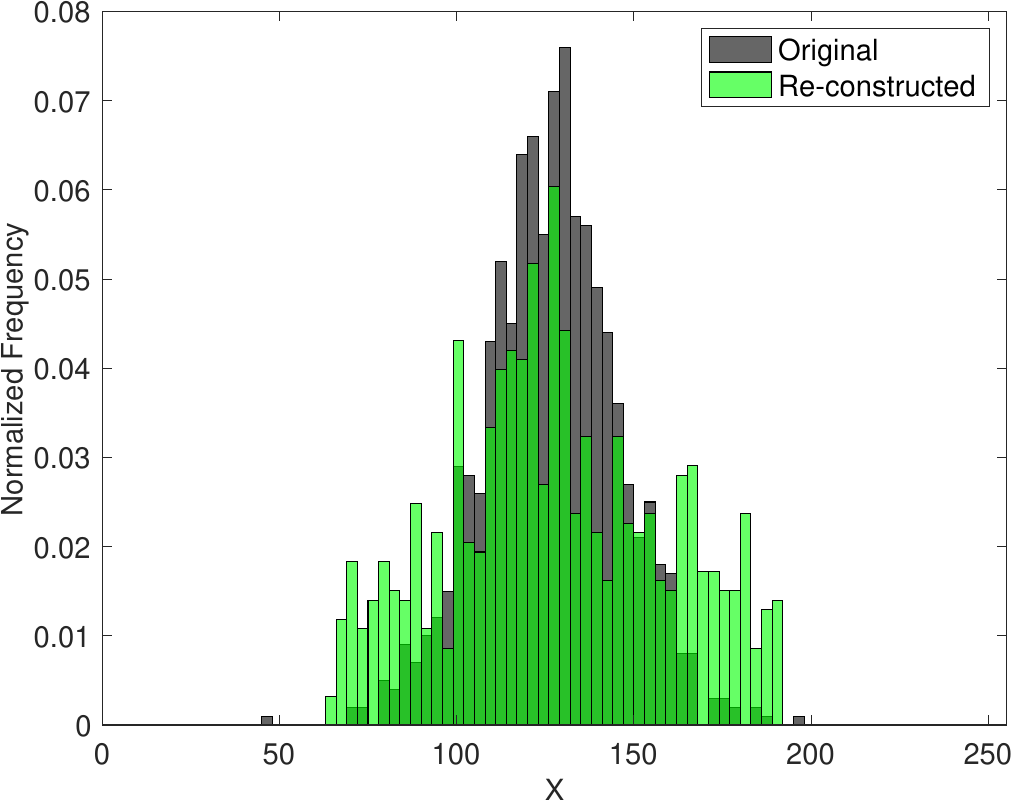}
  \caption{\footnotesize EM-based Re-construction on \textbf{F2}.} \label{em_fp2}
\end{subfigure}\hfill
\begin{subfigure}[t]{0.3\textwidth}
  \includegraphics[width=\linewidth]{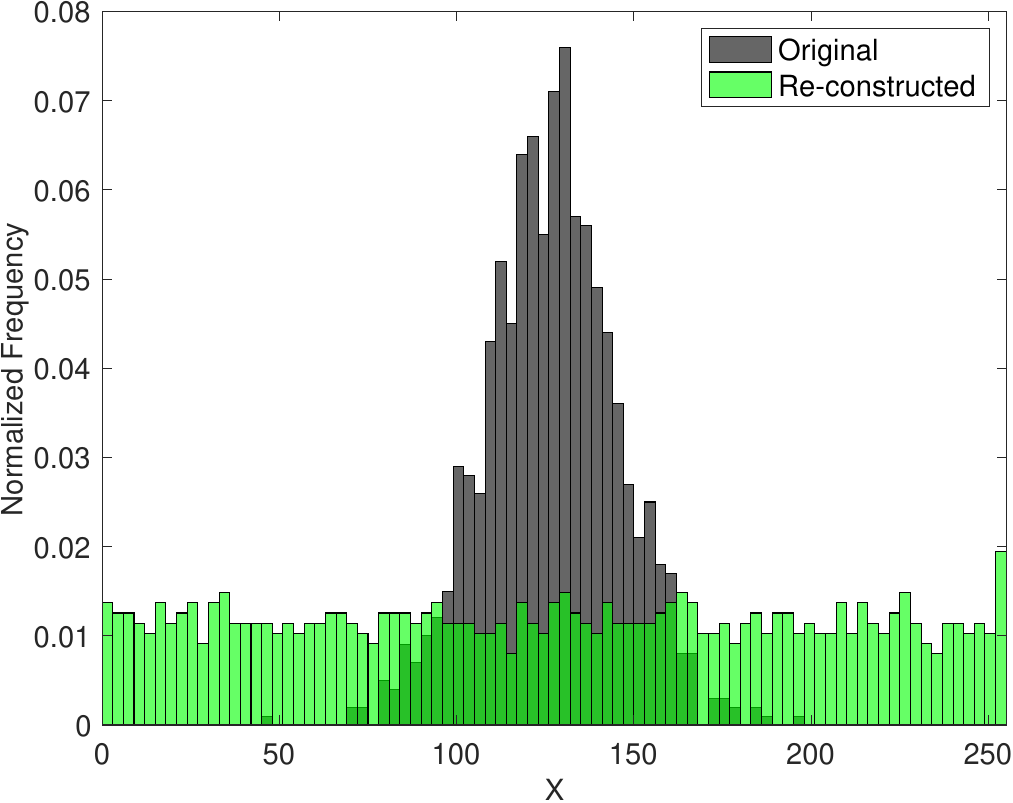}
  \caption{\footnotesize EM-based Re-construction on \textbf{F3}.} \label{em_fp3}
\end{subfigure}\hfill
\\
\vspace{0.05in}
\begin{subfigure}[t]{0.3\textwidth}
  \includegraphics[width=\linewidth]{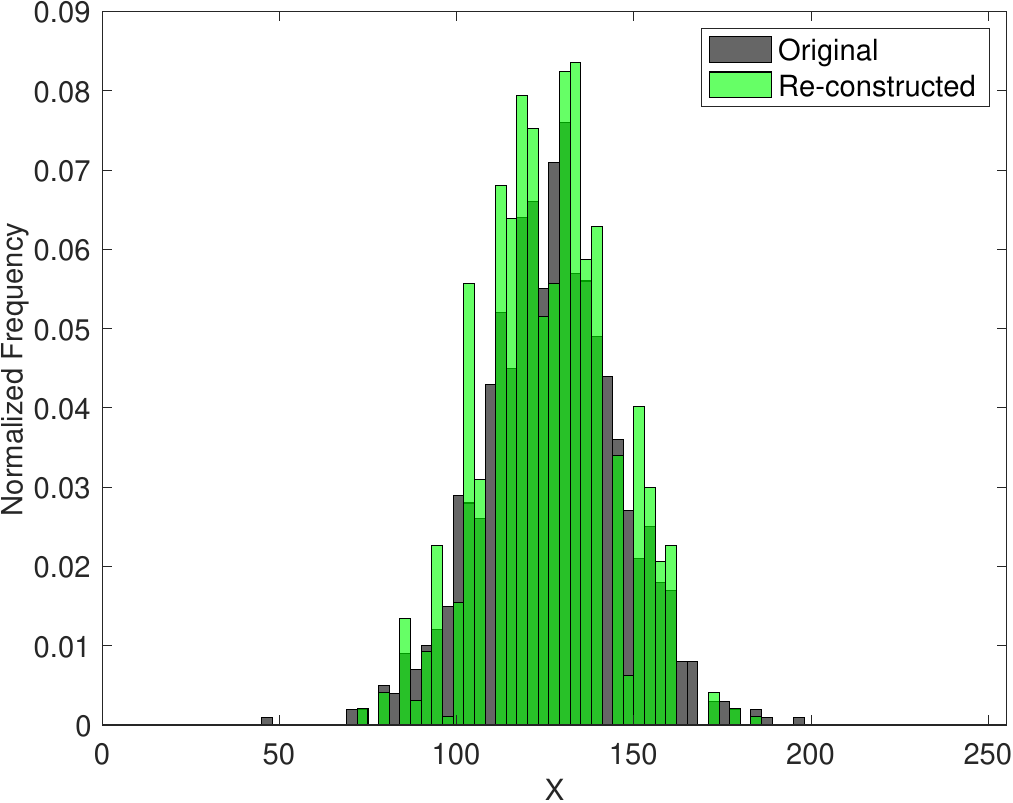}
  \caption{\footnotesize CLR-based Re-construction on \textbf{F1}.} \label{clr_fp1}
\end{subfigure}\hfill
\begin{subfigure}[t]{0.3\textwidth}
  \includegraphics[width=\linewidth]{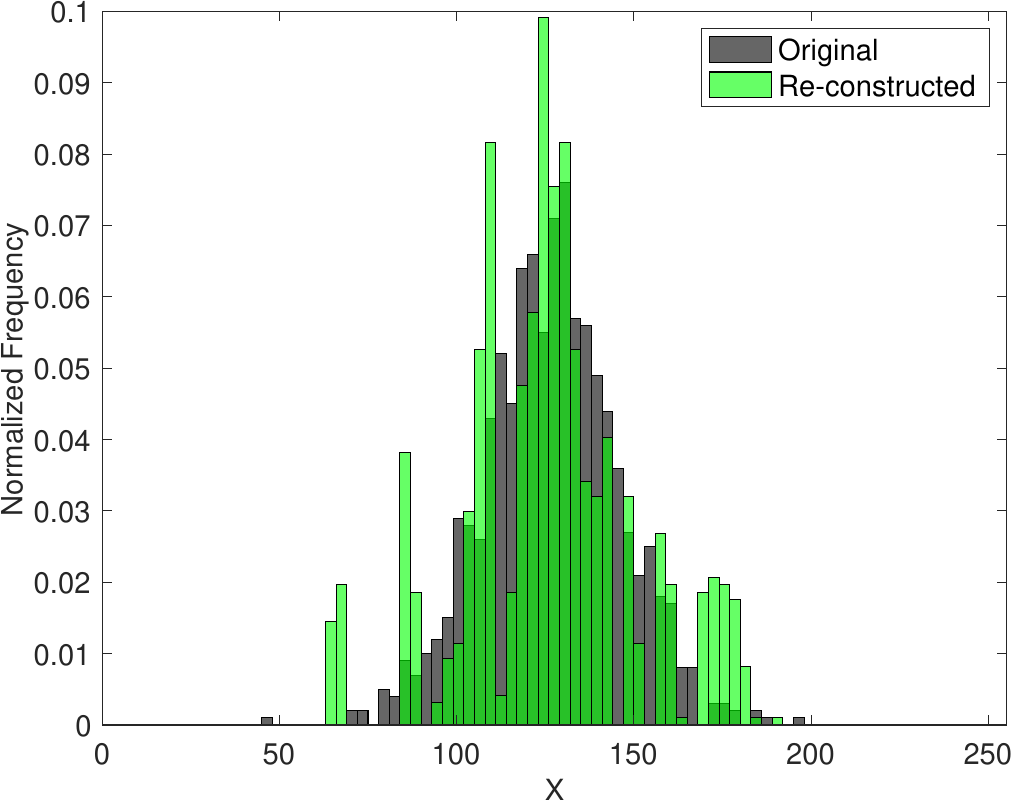}
  \caption{\footnotesize CLR-based Re-construction on \textbf{F2}.} \label{clr_fp2}
\end{subfigure}\hfill
\begin{subfigure}[t]{0.3\textwidth}
  \includegraphics[width=\linewidth]{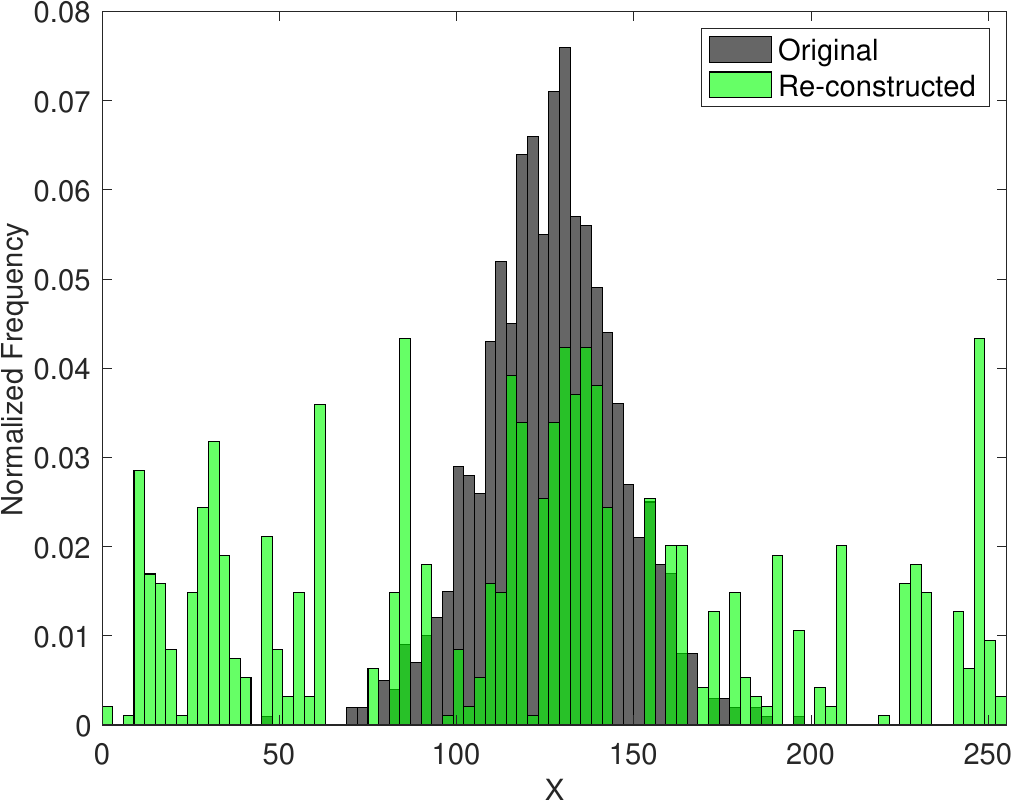}
  \caption{\footnotesize CLR-based Re-construction on \textbf{F3}.} \label{clr_fp3}
\end{subfigure}\hfill
   \caption{\small The normalized frequencies of input data are plotted in grey. Software-generated DP noises are drawn based on three failure patterns \textbf{F1}: $f = [ 0, 0, 0, 0, 0, 0.82, 0.82, 0.82 ]$, \textbf{F2}: $f = [0, 0, 0.82, 0.82, 0.82, 0, 0, 0]$ and \textbf{F3}: $f = [0.82, 0.82, 0.82, 0, 0, 0, 0, 0]$ for $\epsilon = \text{ln}3$. Green plots show the re-constructed distributions under EM-based and CLR-based algorithms.} \label{sim_reconstruction}
\end{figure*}

When it comes to different adversarial knowledge, we evaluate how it can affect adversaries' capability in compromising data privacy. We present simulation results in Figure~\ref{fig:ia}. When compared with an adversary with \textbf{K1}, its peer with \textbf{K2} outperforms in inference accuracy at the high end of the spectrum (i.e., failed cells close to MSB). This can be reasoned by looking into the characteristics of the input data. As shown in Figure~\ref{sim_reconstruction}, there are nearly no samples beyond $X = 192$ (in binary form 110*****), so the \textbf{K2} can confidently eliminate 110***** and 111***** from the \emph{indistinguishable set} leading to higher inference accuracy. Nevertheless, we notice that the \textbf{K1} adversary does not have an evident advantage over \textbf{K2} when failed cells are clustered at low-to-medium bit positions. This provides us with a great insight when rendering cell failures at the design of \texttt{SRAM\_DP}.

\begin{figure}[!t]
\begin{subfigure}[t]{0.235\textwidth}
  \includegraphics[width=\linewidth]{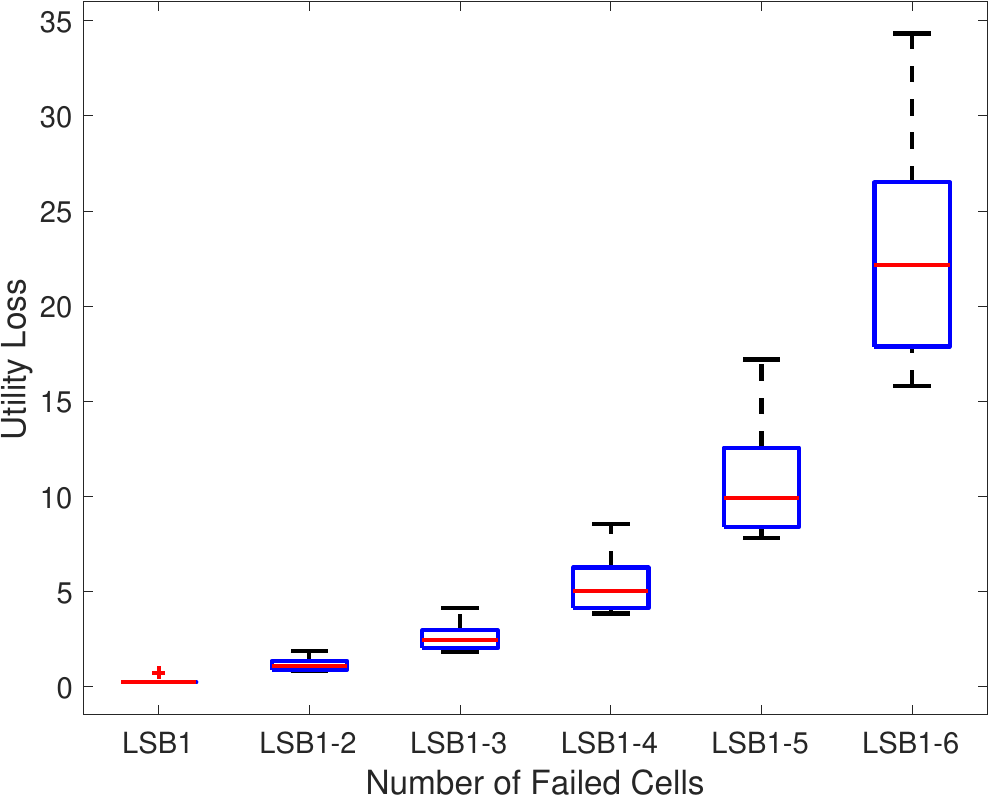}
  \caption{\small UL w.r.t. different z} \label{ul_num}
\end{subfigure}
\begin{subfigure}[t]{0.235\textwidth}
  \includegraphics[width=\linewidth]{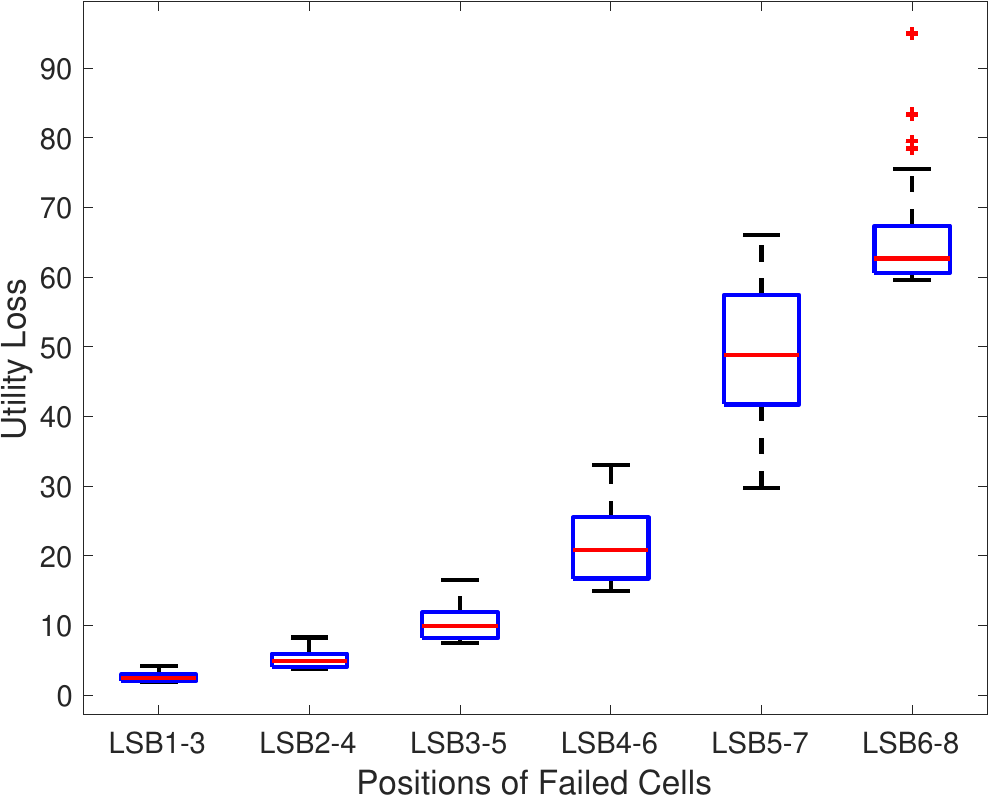}
  \caption{\small UL w.r.t. cell failure positions} \label{fig:ul}
\end{subfigure}
   \caption{Utility analysis.} \label{fig:utility}
\end{figure}

\textbf{Utility Analysis:} In addition to the privacy analysis, we evaluate utility losses. In this simulation, we set $\epsilon = \text{ln}3$. As shown in Figure~\ref{ul_num}, it is obvious that utility loss increases exponentially as there are more failed cells. This finding matches with our theoretical analysis in Theorem \ref{DP2}. Nevertheless, it is not advisable to select a small number of failed cells for the reasons that were discussed in the privacy analysis. 

Moreover, Figure~\ref{fig:ul} shows that the utility loss is moderately low when failed cells are placed at the low end of bit positions, which is intuitive.

\textbf{Data Re-construction:} Here, we demonstrate how well original data can be re-constructed from the sanitized one by using EM-based and CLR-based algorithms. Specifically, in the EM-based algorithm, we set the convergence criteria of as $\delta = 10^{-3}$ and assume no prior information about input statistics. Likewise, no information about original distribution is assumed in the CLR-based algorithm. We consider three failed cells with failure patterns \textbf{F1}, \textbf{F2} and \textbf{F3} for $\epsilon = \text{ln}3$. Figure~\ref{sim_reconstruction} shows the reconstructed frequency histograms versus the original one. There are two take-away from the results. First, both algorithms in general perform better when the failed positions are near LSB. Second, the EM-based algorithm is preferred at low-end bit failures while the CLR-based one slightly outperforms its peer at high-end bit failures. Another note is that if any prior information were available such as variance, we could significantly improve upon reconstructing the shape of the distribution via smoothing.

\subsection{Experiment Results and Analysis}
\vspace{0.05in}
A key take-away from the above simulation results is that the number of, the location of, and the failure rate of failed cells jointly affect IA, UL, and data re-construction accuracy. A fair principle is to place failed cells as close to LSBs as possible and keep the number of failed cells moderate. In this subsection, we instantiate a memory configuration for implementation that balances IA, UL, and data re-construction accuracy. That is, we set four failed cells at the four LSBs (i.e., LSB1-4) with the same failure rate whose value is guided by Theorem \ref{DP}. 

The design details of \texttt{SRAM\_DP} are outlined in the Appendix B. In brief, \texttt{SRAM\_DP} is implemented based on a 45 nm CMOS technology and the standard supply voltage is 1V. \texttt{SRAM\_DP} is designed with a layout of 8 memory banks and each bank has 128 words $\times$ 10 bits. 

\textbf{Analysis on IA, UL and Data Re-construction:} Our experiment is based on the setting of 0.50V supply voltage. We write and then read out the same 1,000 data, and analyze the performance of IA, UL and data re-construction compared with the previous simulation results. 

\begin{figure}[!t]
\begin{subfigure}[t]{0.235\textwidth}
  \includegraphics[width=\linewidth]{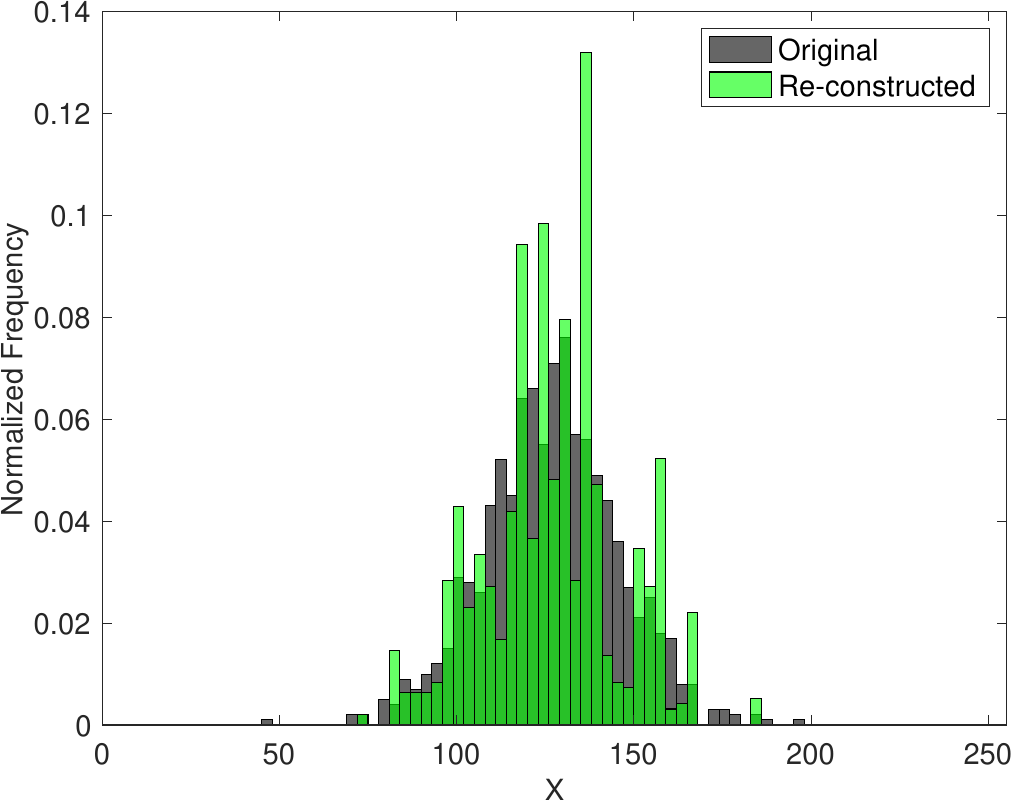}
    \caption{\small Reconstructed data distribution for $\epsilon = 1.49$.}
  \label{fig:exp:em}
\end{subfigure}
\begin{subfigure}[t]{0.235\textwidth}
  \includegraphics[width=\linewidth]{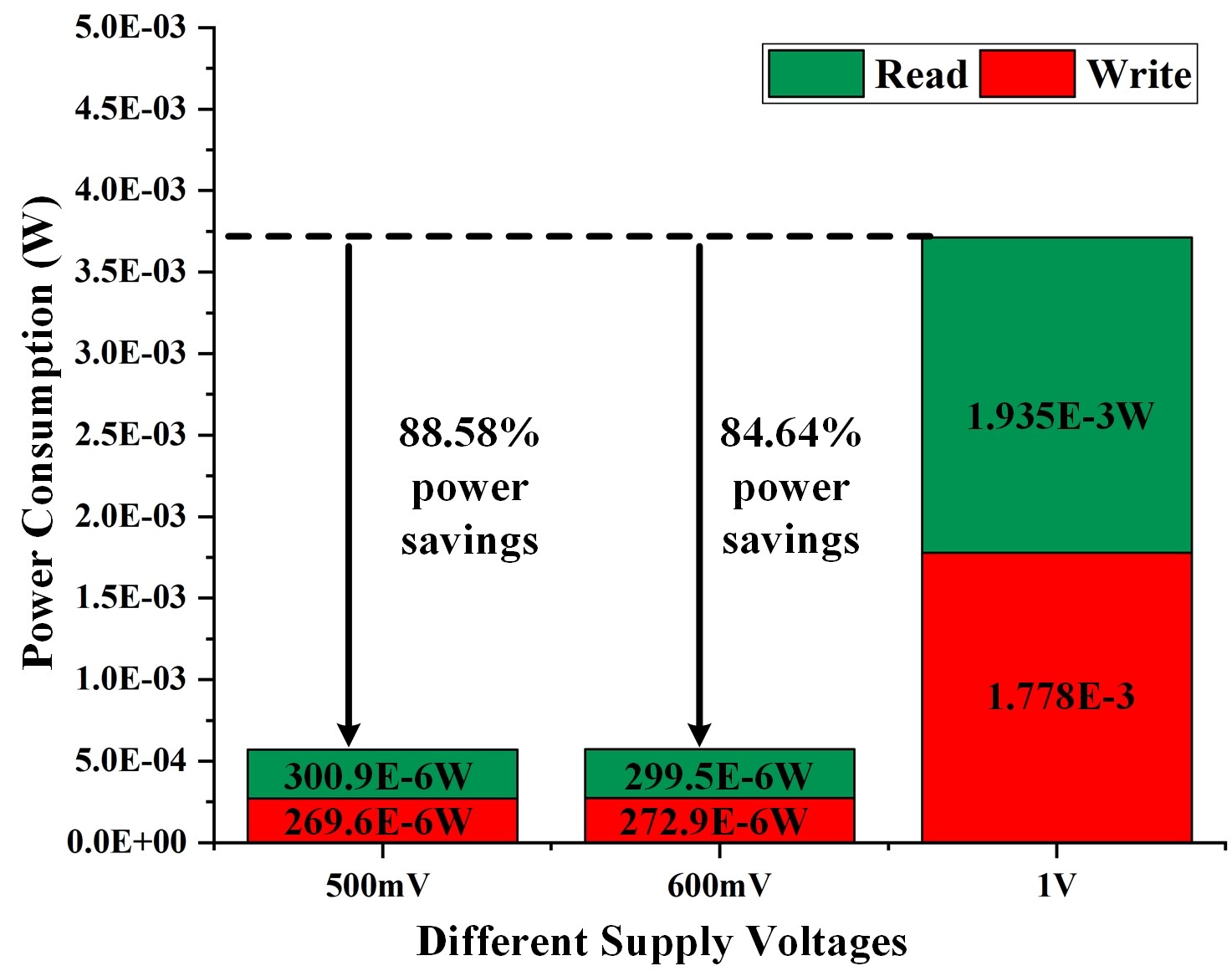}
  \caption{Power savings.}
  \label{fig:power}
\end{subfigure}
   \caption{Estimation accuracy and power saving.}
\end{figure}

First, as shown in Figure~\ref{fig:exp:em}, the re-constructed data is plotted against the original data. Obviously, the EM-based algorithm has a quite low miss-detection rate and its frequency estimation for most data records are accurate. Nonetheless, if we use the Mean Squared Error (MSE) metric to capture how well data is reconstructed, the EM-based algorithm achieves MSE = 69.56 in the experiment while MSE = 5.12 in the simulation. Apparently, the experimental performance is inferior to the simulation. The reason is because of the inconsistency of failure rate $f$ in the sense that EM-based algorithm takes the simulated or theoretical value of $f$ (i.e., 81.57\%) as the input parameter to recover the experimentally perturbed data, in which the exact value of $f$ may deviate from the simulated one.

\begin{figure}[!t]
\begin{subfigure}[t]{0.23\textwidth}
  \includegraphics[width=\linewidth]{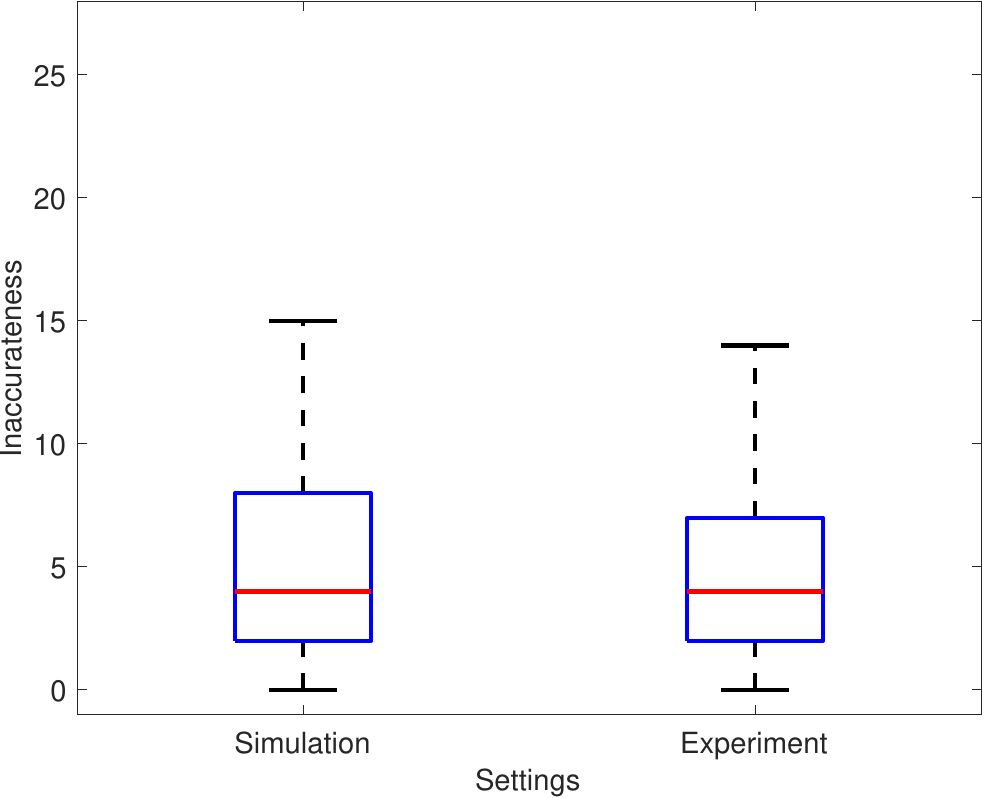}
  \caption{\small IA comparison with \textbf{K1}} \label{exp_ia}
\end{subfigure}
\begin{subfigure}[t]{0.23\textwidth}
  \includegraphics[width=\linewidth]{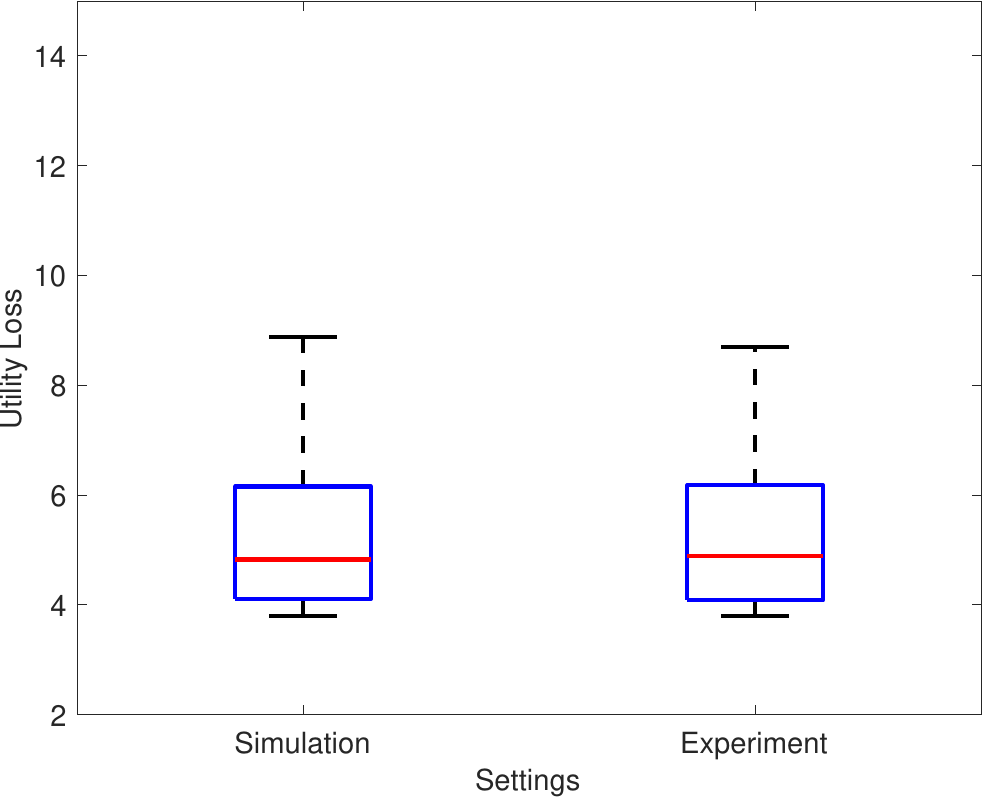}
  \caption{\small UL comparison} \label{exp_ul}
\end{subfigure}
   \caption{Performance Comparison on IA and UL.} \label{fig:exp:ia_ul}
\end{figure}

Next, we demonstrate and compare the performance of IA and UL. As shown in Figure~\ref{fig:exp:ia_ul}, the simulation and experiment results are comparatively similar, which implies that the perturbed data that is generated in the experiment resembles well to the one generated in the simulation. 

\textbf{Power Saving:} Figure~\ref{fig:power} shows the power consumption of \texttt{SRAM\_DP} at three different voltages including standard supply voltage 1V and two low voltages 0.60V and 0.50V. For each test case shown in Figure~\ref{fig:power}, the average power consumption is measured for writing 8-input input data 10100110\textsubscript{2} to a random word in a 128 word $\times$ 10 bit memory bank, which is initialized to 10101001\textsubscript{2}, followed immediately by reading 10100110\textsubscript{2} from the same word, such that all read/write memory operations are equally included (i.e., reading “0” and “1,” and writing “0” to “0,” “0” to “1,” “1” to “0,” and “1” to “1”). During this process, the two MSBs of the selected 10-bit word store the pattern selection signal S[1:0], which is generated by the 2-bit random generator (Figure~\ref{fig:random}). Our first test case is that \texttt{SRAM\_DP} operates at 1V, which is the standard supply voltage without memory failures, and it shows that the total power consumption is $3.713 \times 10^{-3}$W. Our next test case is to let \texttt{SRAM\_DP} operate at 0.60V, the cell failure rate is 60.26\%, and the corresponding $\epsilon$ value is 3.36 (Table \ref{tab:fail:dp}). With noise injection, \texttt{SRAM\_DP} consumes $5.724 \times 10^{-4}$W, which is 84.64\% lower than base case. Our final test case is that \texttt{SRAM\_DP} operates at 0.50V and the cell failure rate is increased to 81.57\%. In this case, \texttt{SRAM\_DP} achieves the LDP with $\epsilon$ = 1.49 and it obtains 88.58\% power savings as compared to the base case.

\begin{figure}[!t]
  \includegraphics[width=\linewidth]{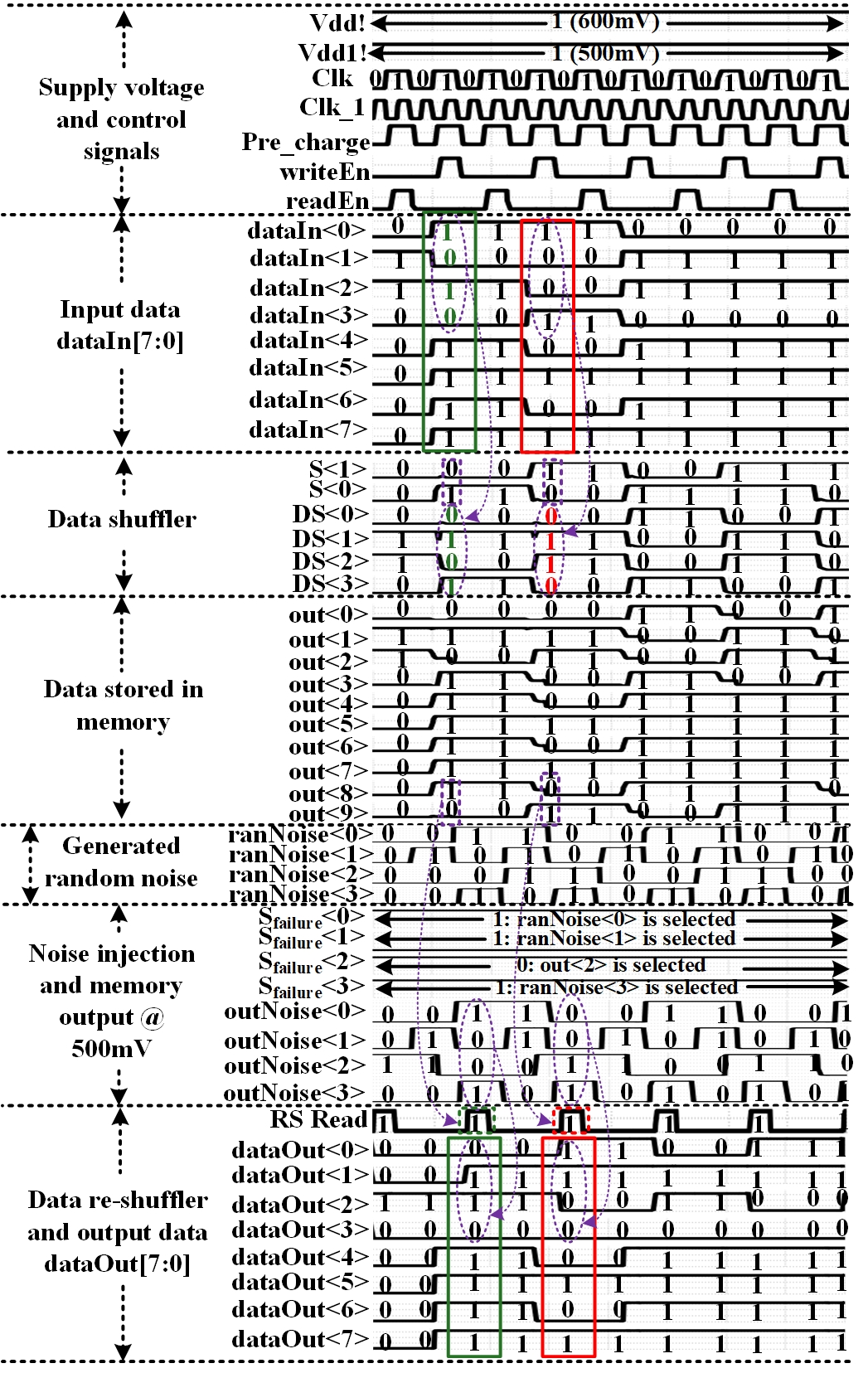}
  \centering
  \caption{Timing diagram of the implemented \texttt{SRAM\_DP}.}
  \label{fig:timing_diagram}
\end{figure}

\textbf{Timing Diagram:} The timing diagram of \texttt{SRAM\_DP} is shown in Figure~\ref{fig:timing_diagram}. Two supply voltages are used in \texttt{SRAM\_DP} including Vdd! and Vdd1!. Vdd1! is applied for memory cells and it can be adjusted to enable the target failure rate for noise injection, as shown in Table \ref{tab:fail:dp}. In this timing diagram, the value of Vdd1! is 0.50V. Vdd! is used as a global supply voltage for all other designs to support the functionality of the memory and its values is 0.60V. Also, two clock signals are adopted in \texttt{SRAM\_DP}: Clk is the global clock signal for the entire memory and Clk\textunderscore1 is used to generate random noise ranNoise[3:0] for four LSBs in the noise injection process. writeEn and readEn are write enable and read enable signals, respectively. dataIn[7:0] is 8-bit input data and dataOut[7:0] is the final output of the memory. The 2-bit control signal S[1:0] are the generated 2-bit random bits for permutation pattern selection. 

The working process is detailed as follows. As one example which is highlighted in green in Figure~\ref{fig:timing_diagram}, 8-bit input data dataIn[7:0] = 11110101\textsubscript{2} is applied to the memory. The generated random signal S[1:0] is 01\textsubscript{2}, so the permutation pattern $\pi_{2}$-[0, 1, 2, 3, 5, 4, 7, 6] is selected for data shuffling and the four LSBs DS[3:0] are reordered to 1010\textsubscript{2}. Then, 10-bit out[9:0] including 8-bit shuffled data DS[7:0] and 2-bit pattern selection signal S[1:0] are written to memory for storage. During the reading process, the generated random noise ranNoise will be injected according to the memory failure map at 0.50V. For the selected word, three bits out of the four LSBs are failed and the control signal S\textsubscript{failure}[3:0] are generated accordingly. After noise injection, the value of four LSBs outNoise[3:0] is 1001\textsubscript{2}. Then, based on the control signal out[9:8], i.e., S[1:0], the data re-shuffler is enabled to reverse the bit shift and generate the final output dataOut[7:0], which is 11110110\textsubscript{2}. As another example highlighted in red in Figure~\ref{fig:timing_diagram}, the 8-bit input data dataIn[7:0] is 10101001\textsubscript{2} and the pattern selection signal S[1:0] is 10\textsubscript{2}, and therefore the permutation pattern $\pi_{3}$-[0, 1, 2, 3, 6, 7, 4, 5] is selected for data shuffling. The generated 10-bit out[9:0] including 8-bit shuffled data 10100110\textsubscript{2} and 2-bit pattern selection signal 10\textsubscript{2} are stored in memory. During the read operation, based on the control signal S\textsubscript{failure}[3:0], ranNoise$\langle 0 \rangle$, ranNoise$\langle 1 \rangle$, out$\langle 2 \rangle$, and ranNoise$\langle 3 \rangle$ are selected as four LSBs of the data (1100\textsubscript{2}). After re-shuffling, the final output data is 10100011\textsubscript{2}.

\subsection{Comparison with Software-based LDP} In this section, we compare our \texttt{SRAM\_DP} mechanism with the widely accepted software-based RR mechanisms for their performance in utility, estimation error, and system overhead.

\textbf{Setup:} The comparison is drawn by using a real-life dataset --- \emph{Foursquare} which is one of the largest location datasets \cite{yang2014modeling}. \emph{Foursquare} contains 227,428 check-ins in New York City, each of which is associated with a user ID, time stamp and GPS coordinate. Specifically, we narrow down the geographical scope to a 2.8Km$\times$2.8Km region within (40.7999N, -73.9700W) and (40.7744N, -73.9445W). A total of 4,759 check-ins are found within this region. We further divide this region into 64$\times$64 areas of interests (AOIs) --- each AOI is a 43.75m$\times$43.75m grid. For this selected region, all check-in coordinates (longitude and latitude) are the same to their tenths decimal, so the \texttt{SRAM\_DP} and RR mechanisms are only applied to the hundredths, thousandths and ten thousandths, i.e., **.*999-**.*744 for latitude while **.*700-**.*445 for longitude.

To implement RR, we employ the IBM Diffprvlib \cite{holohan2019diffprivlib} for which we call the \emph{diffprivlib.mechanisms.Binary} class to add LDP noises. To draw a fair comparison, we consider an utility-optimal RR that the bit randomization is only applied to the same LSBs as \texttt{SRAM\_DP}. RR as software-based LDP mechanisms are programmed, compiled and run in the VS Code IDE in a macOS v12.3 computer with a Apple M1 chip of 39-watt standard power and a 32GB memory of 12-watt standard power.

\begin{figure}[!t]
\begin{subfigure}[t]{0.235\textwidth}
  \includegraphics[width=\linewidth]{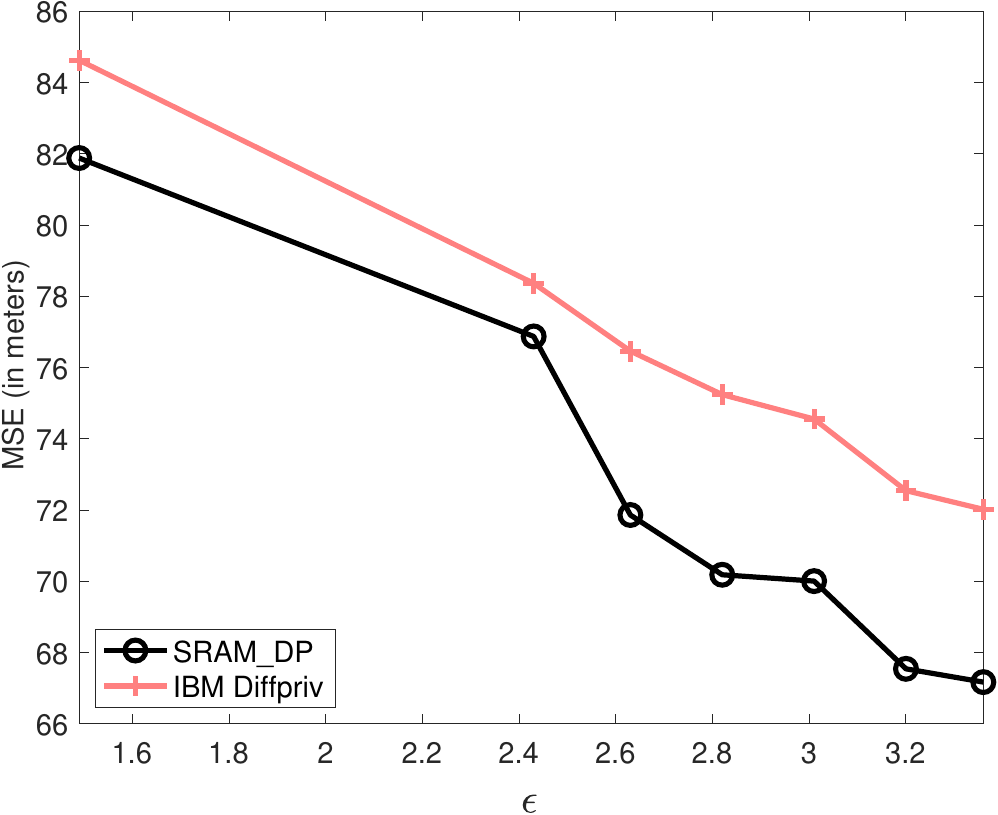}
    \caption{\small Utility Losses.}
  \label{fig:ibm:ul}
\end{subfigure}
\begin{subfigure}[t]{0.235\textwidth}
  \includegraphics[width=\linewidth]{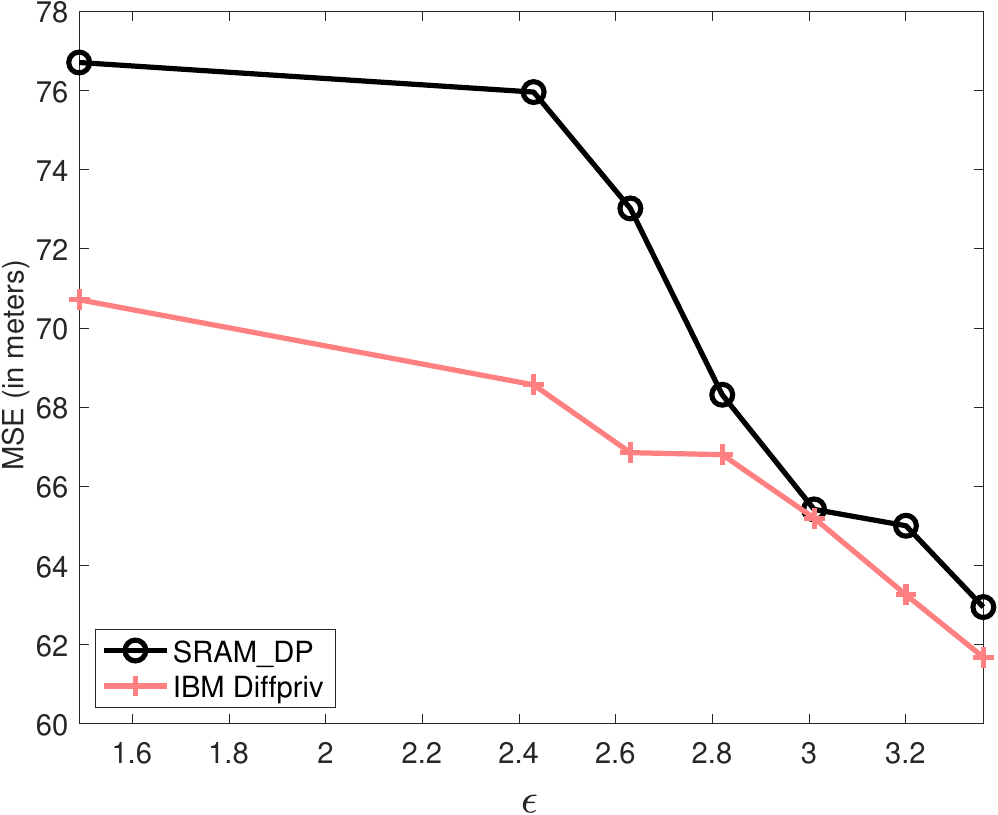}
  \caption{Estimation Errors.}
  \label{fig:ibm:est}
\end{subfigure}
   \caption{Comparison of \texttt{SRAM\_DP} and the utility-optimal RR mechanism from IBM Diffprvlib.}
\end{figure}

\textbf{UL and Estimation Accuracy:} We adopt the metric MSE in meters to assess the utility losses and estimation errors. EM algorithm is employed for estimating original data. As shown in Figure \ref{fig:ibm:ul}, \texttt{SRAM\_DP} achieves lower utility loss by as much as 6.89\%, but the estimation error can be 8.45\% higher than the RR mechanism. The reason lies within the calculation and application of cell failure rate vector $\{f_1,...,f_8\}$. For simplicity, our \texttt{SRAM\_DP} mechanism calculates the average cell failure rate across all 1,000 wordlines and uses it for recovering original data. Yet, each wordline may not exactly follow the failure rate $\{f_1,...,f_8\}$, thus leading to high estimation errors. One possible remedy to this problem is to keep track of each wordline's cell failure rate, but it comes with a sacrifice in storage overhead. Recall that our insights from Figure \ref{fig:exp:em} coincide with our observation here, that is the incapability of tracking $f$s in fine granularity deteriorates estimation accuracy. 

\textbf{System Overhead:} By using the \emph{psutil} tool and excluding the most time-consuming library loading process, we obtained the 8.71\% CPU usage and 0.287\% memory usage for the 8-bit Binary randomization process, which accounts for 7.61$\times 10^{-5}$ seconds runtime and consumes roughly $39 \times 8.71\% + 12 \times 0.287\%$  = 3.43 watt power. 

Moreover, we consider a standard memory (45 nm CMOS, 8 memory banks, each bank has 128 words $\times$ 10 bits) without any customization as the baseline for comparison. As shown in Table \ref{tab:overhead}, despite minor increase in chip peripheral overhead due to added circuits for voltage control and bit manipulation, \texttt{SRAM\_DP} significantly outperforms IBM Diffprvlib in terms of system responsiveness (measured in latency) and power saving. Note that we neglect the storage overhead for the bit-shuffle pattern as it is data-dependent and can be easily optimized using data piggybacking, peripheral look-up table, and many other methods.
\begin{figure}
\centering
 \includegraphics[width=0.75\linewidth]{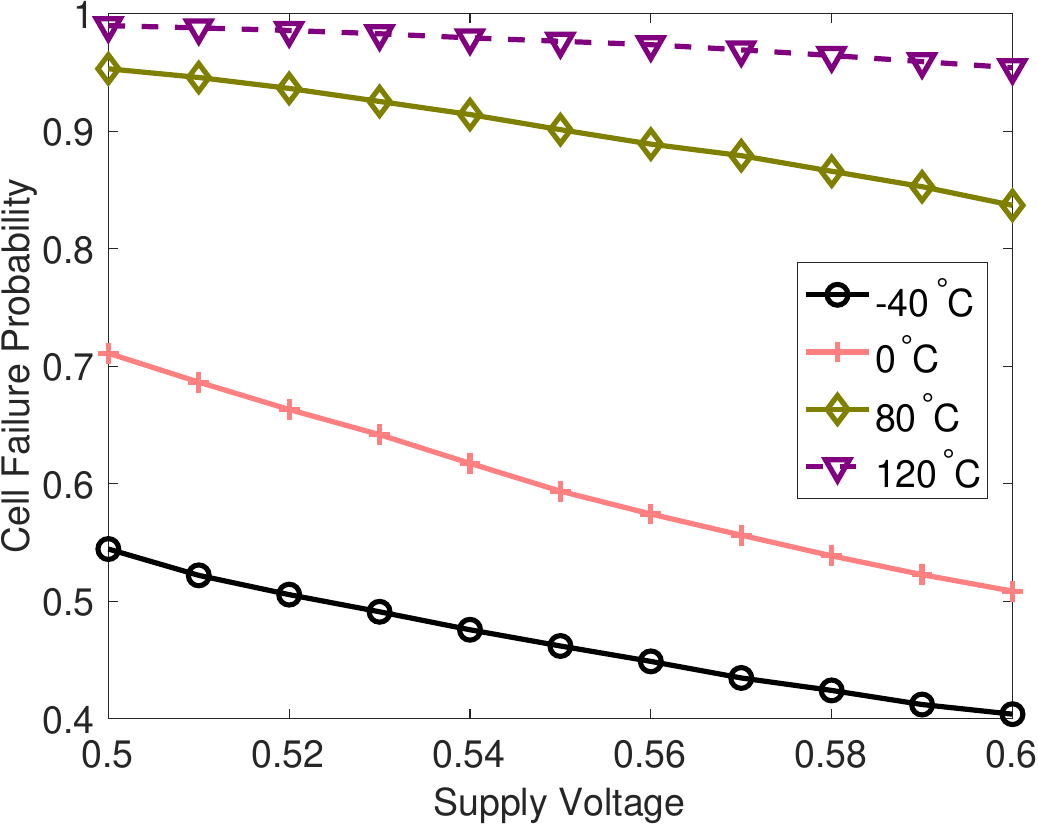}
  \caption{Cell failure probabilities at various temperatures.}
  \label{fig:temperature}
\end{figure}
\begin{table*}[t]
\caption{Performance comparison between \texttt{SRAM\_DP} and utility-optimal RR for $\epsilon = 1.49$ (bold ones are better)}
\centering
\footnotesize
\begin{adjustbox}{width=\textwidth}
 \begin{tabular}{c || c | c | c} 
 \hline
Figure of Merits & Baseline & \texttt{SRAM\_DP} & IBM Diffprivlib \\ 
 \hline\hline
Utility Loss (MSE) & 0 & \textbf{78.8861 (-6.79\%)}  & 84.6308 \\ 
 \hline
Estimation Error (MSE) & 0 & 76.7106  & 
\textbf{70.717 (-7.81\%)} \\ 
 \hline
Chip Peripheral Overhead (\# of transistors) & 66,000 & 67,623 (+2.459\%)  & \textbf{+0\%} \\ 
 \hline
System Latency (ns) & 0.84 & \textbf{1.02 (+21.4\%)} & $7.61\times10^{4}$ (+9.06$\times 10^6$\%) \\
 \hline
Power Consumption (mW) & 3.713 & \textbf{0.5724 (-88.58\%)} & $3.43 \times10^{3}$ (+922.78\%) \\
 \hline
\end{tabular}
\label{tab:overhead}
\end{adjustbox}
\end{table*}

\section{Discussion on \texttt{SRAM\_DP} Reliability}\label{reliability}
\textbf{Temperature Variations}: The reliability of the proposed \texttt{SRAM\_DP} in various operating conditions such as high aging and temperature variation is one important design consideration. Figure~\ref{fig:temperature} shows the simulated failure probability of the 6T cells used in our design under temperature fluctuations. The temperature was varied from -40$^{\circ}$C to 120$^{\circ}$C for the target voltage range (0.5V-0.6V), with 25$^{\circ}$C as the reference temperature. As shown, the memory cells become more inclined to failure as the temperature increases (2 times from -40$^{\circ}$C to 125$^{\circ}$C). This is mainly due to the weakened PMOS transistors in the 6T SRAM cells.  

\textbf{Voltage Precision Control}: Supply voltage droop is another factor which will cause variation of memory failure probability, thereby impacting the reliability of \texttt{SRAM\_DP}. However, its effect is much smaller than the temperature variation for today's mainstream voltage regulators and power management Integrated Circuits (PMIC). For example, the voltage droop of a commercial PMIC for IoT devices can be well controlled within $\pm$ 1.5\% \cite{digikey19}. Such a small voltage droop can cause less than $\pm$ 1\% change in the cell failure probability, as shown in Figure~\ref{fig:failure_rate} and \ref{fig:temperature}. When mapping the cell failure probability variation into the the drift of LDP parameter $\epsilon$, we have the following theorem.
\begin{theorem}
For a SRAM cell whose failure probability changes from $f_i$ to $\alpha_i f_i$ ($0 \leq \alpha_i \leq 1/f_i$) due to the supply voltage droop, the variation of $\epsilon$ is bounded by $|\Delta \epsilon| = |\epsilon^\prime - \epsilon|$ $\leq$ $\sum_{i=1}^{n} |ln(2\alpha_i - 1)|$.
\end{theorem}
\begin{proof}
According to Theorem \ref{DP}, we can calculate that
\begin{align*}
\begin{split}
|\Delta \epsilon| = & |\sum_{i=1}^{n} ln(\frac{1 - \frac{1}{2}\alpha_i f_{i}}{\frac{1}{2}\alpha_i f_{i}}) - \sum_{i=1}^{n}ln(\frac{1 - \frac{1}{2}f_{i}}{\frac{1}{2}f_{i}})|  \\
= & |\sum_{i=1}^{n} ln(2 - \alpha_i f_i) - ln(2 -  f_i) - ln(\alpha_i)| \\ 
\leq & \sum_{i=1}^{n} |ln(2 - \alpha_i f_i) - ln(2 -  f_i)| + \sum_{i=1}^{n} |ln(\alpha_i)| 
\end{split}
\end{align*}
For the case $1 \leq \alpha_i \leq 1/f_i$ implying that the supply voltage droop causes the cell failure probability to increase, we can deduce that for every SRAM cell, $ln(2 - f_i) - ln(2 - \alpha_i f_i)$ $\leq$ $ln (2 - \frac{1}{\alpha_i})$ by observing that $\frac{2 - f_i}{2 - \alpha_i f_i} - (2 - \frac{1}{\alpha_i})$ = $2(\alpha_i - 1)(\alpha_i f_i - 1)$ $\leq 0$. Similarly, for the cell probability downscaling case where $0 \leq \alpha_i \leq 1$, we can arrive at the same upper bound. Then, $|\Delta \epsilon|$ can be further written as 
\begin{align*}
|\Delta \epsilon| \leq \sum_{i=1}^{n} |ln(2 - \frac{1}{\alpha_i})| + \sum_{i=1}^{n} |ln(\alpha_i)| = \sum_{i=1}^{n} |ln(2\alpha_i - 1)|.
\end{align*}
\end{proof}
Specifically for our developed \texttt{SRAM\_DP} that adopts 6T cells and renders cell failures in the 4 LSBs, a $\pm$1\% drift in cell probability will result in at most $\pm$0.08 variations in $\epsilon$ for all possible selection of $\epsilon$'s. The privacy customer should thus be cautioned of this worst-case variation when using the \texttt{SRAM\_DP} chip.

\textbf{Reliability Enhancement Solutions}: Amid the reliability concerns, different design strategies can be adopted as remedies. First, designing variation-aware cells is one possible solution. The current \texttt{SRAM\_DP} adopts 6T cells to achieve the target failure probability for noise injection. To enhance the reliability, the transistor sizing and structure of memory cells can be further custom designed to enable the target mean failure probability while minimizing its variance in the presence of temperature variation or voltage voltage droop \cite{kwon2021variation}.

Also, \texttt{SRAM\_DP} can utilize the well-studied variation-adaptive solutions \cite{bowman2018adaptive}, which integrate with on-chip sensors and adaptive control circuits to measure specific parameters (e.g., temperature, voltage droop, or aging effect) and then to adjust the supply voltage accordingly. Existing variation-adaptive schemes (such as \cite{tschanz2007adaptive}) have been developed mainly to compensate for the impact of the parameter changes on performance or power consumption. However, they can be easily adapted to our \texttt{SRAM\_DP} design, which requires compensation to achieve a stable failure probability (or $\epsilon$). 

\section{Related Work}\label{relate}
Broadly speaking, most recent works on DP focus on developing domain-specific DP mechanisms for various applications \cite{liu2019dpavatar,pei2023privacy,gai2022efficient}. However, little to no attention has been paid to the realization of DP mechanisms due to the presumption that DP noises can be easily, correctly, and even securely generated and injected in software. This is unfortunately challenged by many recent studies \cite{mironov2012significance,andrysco2015subnormal,haeberlen2011differential} and the emerging lightweight IoT devices that have no rich software modules. Ilvento et al. \cite{ilvento2020implementing} is among the very few works to investigate the realization of DP mechanisms, and they switch from base-e to base-2 arithmetic --- offering higher precision in existing arithmetic libraries --- when implementing the popular exponential DP mechanism in software. 

In recent years, we observe a growing trend of realizing DP, or more broadly noise-injection, mechanisms by utilizing hardware characteristics. In general, the hardware-based approach is superior than the software-based one because of its true randomness of noise and scalability in implementation (i.e., independence from complex software modules), although the former approach falls short in real-time adaptability after hardware is fabricated. Nonetheless, there are only limited number of works along this theme line. Yang et al. \cite{yang2017approximate} proposed to introduce DP Gaussian noises to deep learning model training by scaling down supply voltage and creating SRAM bit errors. Fu et al. \cite{fu2020memristor} leveraged the inherent Gaussian noises due to the imperfectness of memoristor operations and developed a differentially private deep learning model. Although these works are innovative in their ideas, they are neither implemented in hardware nor rigorous when it comes to conforming to DP notions. For instance, Gaussian mechanism only ensures ($\epsilon, \delta$)-DP with $\epsilon < 1$, but some works neglected this constraint.

\section{Conclusion}\label{con}
In this paper, we designed, implemented, and evaluated a new memory architecture to achieve local differential privacy in hardware rather than in software. By downscaling supply voltages, LDP noise can be introduced to data, especially those in least significant bits, when it is stored in memory. Compared with existing software-based LDP mechanisms, this paper's analytical, simulated and experimental results all supported the feasibility and superiority of our design in terms of privacy, utility, latency, and system power consumption with moderate compromise in estimation errors and chip overhead. 


\bibliography{ref}

\begin{thebibliography}{10}
\providecommand{\url}[1]{#1}
\csname url@samestyle\endcsname
\providecommand{\newblock}{\relax}
\providecommand{\bibinfo}[2]{#2}
\providecommand{\BIBentrySTDinterwordspacing}{\spaceskip=0pt\relax}
\providecommand{\BIBentryALTinterwordstretchfactor}{4}
\providecommand{\BIBentryALTinterwordspacing}{\spaceskip=\fontdimen2\font plus
\BIBentryALTinterwordstretchfactor\fontdimen3\font minus
  \fontdimen4\font\relax}
\providecommand{\BIBforeignlanguage}[2]{{%
\expandafter\ifx\csname l@#1\endcsname\relax
\typeout{** WARNING: IEEEtran.bst: No hyphenation pattern has been}%
\typeout{** loaded for the language `#1'. Using the pattern for}%
\typeout{** the default language instead.}%
\else
\language=\csname l@#1\endcsname
\fi
#2}}
\providecommand{\BIBdecl}{\relax}
\BIBdecl

\bibitem{zhou2008large}
Y.~Zhou, D.~Wilkinson, R.~Schreiber, and R.~Pan, ``Large-scale parallel
  collaborative filtering for the netflix prize,'' in \emph{International
  conference on algorithmic applications in management}.\hskip 1em plus 0.5em
  minus 0.4em\relax Springer, 2008, pp. 337--348.

\bibitem{dwork2006calibrating}
C.~Dwork, F.~McSherry, K.~Nissim, and A.~Smith, ``Calibrating noise to
  sensitivity in private data analysis,'' in \emph{Theory of cryptography
  conference}.\hskip 1em plus 0.5em minus 0.4em\relax Springer, 2006, pp.
  265--284.

\bibitem{machanavajjhala2008privacy}
A.~Machanavajjhala, D.~Kifer, J.~Abowd, J.~Gehrke, and L.~Vilhuber, ``Privacy:
  Theory meets practice on the map,'' in \emph{Proceedings of the 2008 IEEE
  24th International Conference on Data Engineering}.\hskip 1em plus 0.5em
  minus 0.4em\relax IEEE Computer Society, 2008, pp. 277--286.

\bibitem{mcsherry2009differentially}
F.~McSherry and I.~Mironov, ``Differentially private recommender systems:
  Building privacy into the netflix prize contenders,'' in \emph{Proceedings of
  the 15th ACM SIGKDD international conference on Knowledge discovery and data
  mining}.\hskip 1em plus 0.5em minus 0.4em\relax ACM, 2009, pp. 627--636.

\bibitem{qin2016heavy}
Z.~Qin, Y.~Yang, T.~Yu, I.~Khalil, X.~Xiao, and K.~Ren, ``Heavy hitter
  estimation over set-valued data with local differential privacy,'' in
  \emph{Proceedings of the 2016 ACM SIGSAC Conference on Computer and
  Communications Security}.\hskip 1em plus 0.5em minus 0.4em\relax ACM, 2016,
  pp. 192--203.

\bibitem{erlingsson2014rappor}
{\'U}.~Erlingsson, V.~Pihur, and A.~Korolova, ``Rappor: Randomized aggregatable
  privacy-preserving ordinal response,'' in \emph{Proceedings of the 2014 ACM
  SIGSAC conference on computer and communications security}.\hskip 1em plus
  0.5em minus 0.4em\relax ACM, 2014, pp. 1054--1067.

\bibitem{mironov2012significance}
I.~Mironov, ``On significance of the least significant bits for differential
  privacy,'' in \emph{Proceedings of the 2012 ACM conference on Computer and
  communications security}.\hskip 1em plus 0.5em minus 0.4em\relax ACM, 2012,
  pp. 650--661.

\bibitem{andrysco2015subnormal}
M.~Andrysco, D.~Kohlbrenner, K.~Mowery, R.~Jhala, S.~Lerner, and H.~Shacham,
  ``On subnormal floating point and abnormal timing,'' in \emph{2015 IEEE
  Symposium on Security and Privacy}.\hskip 1em plus 0.5em minus 0.4em\relax
  IEEE, 2015, pp. 623--639.

\bibitem{haeberlen2011differential}
A.~Haeberlen, B.~C. Pierce, and A.~Narayan, ``Differential privacy under
  fire.'' in \emph{USENIX Security Symposium}, 2011.

\bibitem{liu2023privacy}
J.~Liu and N.~Gong, ``Privacy by memory design: Visions and open problems,''
  \emph{IEEE Micro}, vol.~44, no.~1, pp. 49--58, 2024.

\bibitem{CIP1}
S.~Y. Yu, H.~Jiang, S.~Huang, X.~Peng, and A.~Lu, ``Compute-in-memory chips for
  deep learning: Recent trends and prospects,'' \emph{IEEE Circuits and Systems
  Magazine}, vol.~21, no.~3, pp. 31--56, 2021.

\bibitem{warner1965randomized}
S.~L. Warner, ``Randomized response: A survey technique for eliminating evasive
  answer bias,'' \emph{Journal of the American Statistical Association},
  vol.~60, no. 309, pp. 63--69, 1965.

\bibitem{fu2020memristor}
J.~Fu, Z.~Liao, J.~Liu, S.~C. Smith, and J.~Wang, ``Memristor-based
  variation-enabled differentially private learning systems for edge computing
  in iot,'' \emph{IEEE Internet of Things Journal}, vol.~8, no.~12, pp.
  9672--9682, 2020.

\bibitem{Xu2020Model}
Y.~Xu, H.~Das, Y.~Gong, and N.~Gong, ``On mathematical models of optimal video
  memory design,'' \emph{IEEE Transactions on Circuits and Systems for Video
  Technology}, vol.~30, no.~1, pp. 256--266, 2019.

\bibitem{Gong2012Cell}
N.~Gong, S.~Jiang, J.~Wang, B.~Aravamudhan, K.~Sekar, and R.~Sridhar,
  ``Hybrid-cell register files design for improving nbti reliability,''
  \emph{Microelectronics Reliability}, vol.~52, no. 9-10, pp. 1865--1869, 2012.

\bibitem{croon2004physical}
J.~Croon, S.~Decoutere, W.~Sansen, and H.~Maes, ``Physical modeling and
  prediction of the matching properties of mosfets,'' in \emph{Proceedings of
  the 30th European Solid-State Circuits Conference (IEEE Cat. No.
  04EX850)}.\hskip 1em plus 0.5em minus 0.4em\relax IEEE, 2004, pp. 193--196.

\bibitem{Das2021ECC}
H.~Das, A.~A. Haidous, S.~C. Smith, and N.~Gong, ``Flexible low-cost
  power-efficient video memory with ecc-adaptation,'' \emph{IEEE Transactions
  on Very Large Scale Integration (VLSI) Systems}, vol.~29, no.~10, pp.
  1693--1706, 2021.

\bibitem{GOTTSCHO2015fault}
M.~Gottscho, A.~BanaiyanMofrad, N.~D. Dutt, A.~Nicolau, and P.~Gupta, ``Dpcs:
  Dynamic power/capacity scaling for sram caches in the nanoscale era,''
  \emph{ACM Transactions on Architecture and Code Optimization}, vol.~12,
  no.~3, pp. 1--27, 2015.

\bibitem{edstrom2017data}
J.~Edstrom, D.~Chen, Y.~Gong, J.~Wang, and N.~Gong, ``Data-pattern enabled
  self-recovery low-power storage system for big video data,'' \emph{IEEE
  Transactions on Big Data}, vol.~5, no.~1, pp. 95--105, 2017.

\bibitem{wang2022hertzbleed}
Y.~Wang, R.~Paccagnella, E.~T. He, H.~Shacham, C.~W. Fletcher, and
  D.~Kohlbrenner, ``Hertzbleed: Turning power $\{$Side-Channel$\}$ attacks into
  remote timing attacks on x86,'' in \emph{31st USENIX Security Symposium
  (USENIX Security 22)}, 2022, pp. 679--697.

\bibitem{dipta2022df}
D.~R. Dipta and B.~Gulmezoglu, ``Df-sca: dynamic frequency side channel attacks
  are practical,'' in \emph{Proceedings of the 38th Annual Computer Security
  Applications Conference}, 2022, pp. 841--853.

\bibitem{murakami2019utility}
T.~Murakami and Y.~Kawamoto, ``Utility-optimized local differential privacy
  mechanisms for distribution estimation,'' in \emph{28th $\{$USENIX$\}$
  Security Symposium ($\{$USENIX$\}$ Security 19)}, 2019, pp. 1877--1894.

\bibitem{wang2017locally}
T.~Wang, J.~Blocki, N.~Li, and S.~Jha, ``Locally differentially private
  protocols for frequency estimation,'' in \emph{26th $\{$USENIX$\}$ Security
  Symposium ($\{$USENIX$\}$ Security 17)}, 2017, pp. 729--745.

\bibitem{yang2014modeling}
D.~Yang, D.~Zhang, V.~W. Zheng, and Z.~Yu, ``Modeling user activity preference
  by leveraging user spatial temporal characteristics in lbsns,'' \emph{IEEE
  Transactions on Systems, Man, and Cybernetics: Systems}, vol.~45, no.~1, pp.
  129--142, 2015.

\bibitem{holohan2019diffprivlib}
N.~Holohan, S.~Braghin, P.~Mac~Aonghusa, and K.~Levacher, ``Diffprivlib: the
  ibm differential privacy library,'' \emph{arXiv preprint arXiv:1907.02444},
  2019.

\bibitem{digikey19}
\BIBentryALTinterwordspacing
Digi-Key, ``Use advanced ldos to meet iot wireless sensor power supply design
  challenges,'' 2019. [Online]. Available:
  \url{https://www.digikey.com/en/articles/use-advanced-ldos-iot-wireless-sensor-power-supply-design}
\BIBentrySTDinterwordspacing

\bibitem{kwon2021variation}
H.~Kwon, D.~Kim, Y.~H. Kim, and S.~Kang, ``Variation-aware sram cell
  optimization using deep neural network-based sensitivity analysis,''
  \emph{IEEE Transactions on Circuits and Systems I: Regular Papers}, vol.~68,
  no.~4, pp. 1567--1577, 2021.

\bibitem{bowman2018adaptive}
K.~A. Bowman, ``Adaptive and resilient circuits: A tutorial on improving
  processor performance, energy efficiency, and yield via dynamic variation,''
  \emph{IEEE Solid-State Circuits Magazine}, vol.~10, no.~3, pp. 16--25, 2018.

\bibitem{tschanz2007adaptive}
J.~Tschanz, N.~S. Kim, S.~Dighe, J.~Howard, G.~Ruhl, S.~Vangal, S.~Narendra,
  Y.~Hoskote, H.~Wilson, C.~Lam \emph{et~al.}, ``Adaptive frequency and biasing
  techniques for tolerance to dynamic temperature-voltage variations and
  aging,'' in \emph{2007 IEEE International Solid-State Circuits Conference.
  Digest of Technical Papers}.\hskip 1em plus 0.5em minus 0.4em\relax IEEE,
  2007, pp. 292--604.

\bibitem{liu2019dpavatar}
J.~Liu, C.~Zhang, B.~Lorenzo, and Y.~Fang, ``Dpavatar: A real-time location
  protection framework for incumbent users in cognitive radio networks,''
  \emph{IEEE Transactions on Mobile Computing}, vol.~19, no.~3, pp. 552--565,
  2019.

\bibitem{pei2023privacy}
X.~Pei, X.~Deng, S.~Tian, J.~Liu, and K.~Xue, ``Privacy-enhanced graph neural
  network for decentralized local graphs,'' \emph{IEEE Transactions on
  Information Forensics and Security}, vol.~19, pp. 1614--1629, 2023.

\bibitem{gai2022efficient}
N.~Gai, K.~Xue, B.~Zhu, J.~Yang, J.~Liu, and D.~He, ``An efficient data
  aggregation scheme with local differential privacy in smart grid,''
  \emph{Digital Communications and Networks}, vol.~8, no.~3, pp. 333--342,
  2022.

\bibitem{ilvento2020implementing}
C.~Ilvento, ``Implementing the exponential mechanism with base-2 differential
  privacy,'' in \emph{Proceedings of the 2020 ACM SIGSAC Conference on Computer
  and Communications Security}, 2020, pp. 717--742.

\bibitem{yang2017approximate}
L.~Yang and B.~Murmann, ``Approximate sram for energy-efficient,
  privacy-preserving convolutional neural networks,'' in \emph{2017 IEEE
  Computer Society Annual Symposium on VLSI (ISVLSI)}.\hskip 1em plus 0.5em
  minus 0.4em\relax IEEE, 2017, pp. 689--694.

\bibitem{hsu2014differential}
J.~Hsu, M.~Gaboardi, A.~Haeberlen, S.~Khanna, A.~Narayan, B.~C. Pierce, and
  A.~Roth, ``Differential privacy: An economic method for choosing epsilon,''
  in \emph{2014 IEEE 27th Computer Security Foundations Symposium}.\hskip 1em
  plus 0.5em minus 0.4em\relax IEEE, 2014, pp. 398--410.

\bibitem{clark2018sram}
L.~T. Clark, S.~B. Medapuram, and D.~K. Kadiyala, ``Sram circuits for true
  random number generation using intrinsic bit instability,'' \emph{IEEE
  Transactions on Very Large Scale Integration (VLSI) Systems}, vol.~26,
  no.~10, pp. 2027--2037, 2018.

\end{thebibliography}
\bibliographystyle{IEEEtran}
\begin{IEEEbiography}
[{\includegraphics[width=1in,height=1.25in,clip,keepaspectratio]{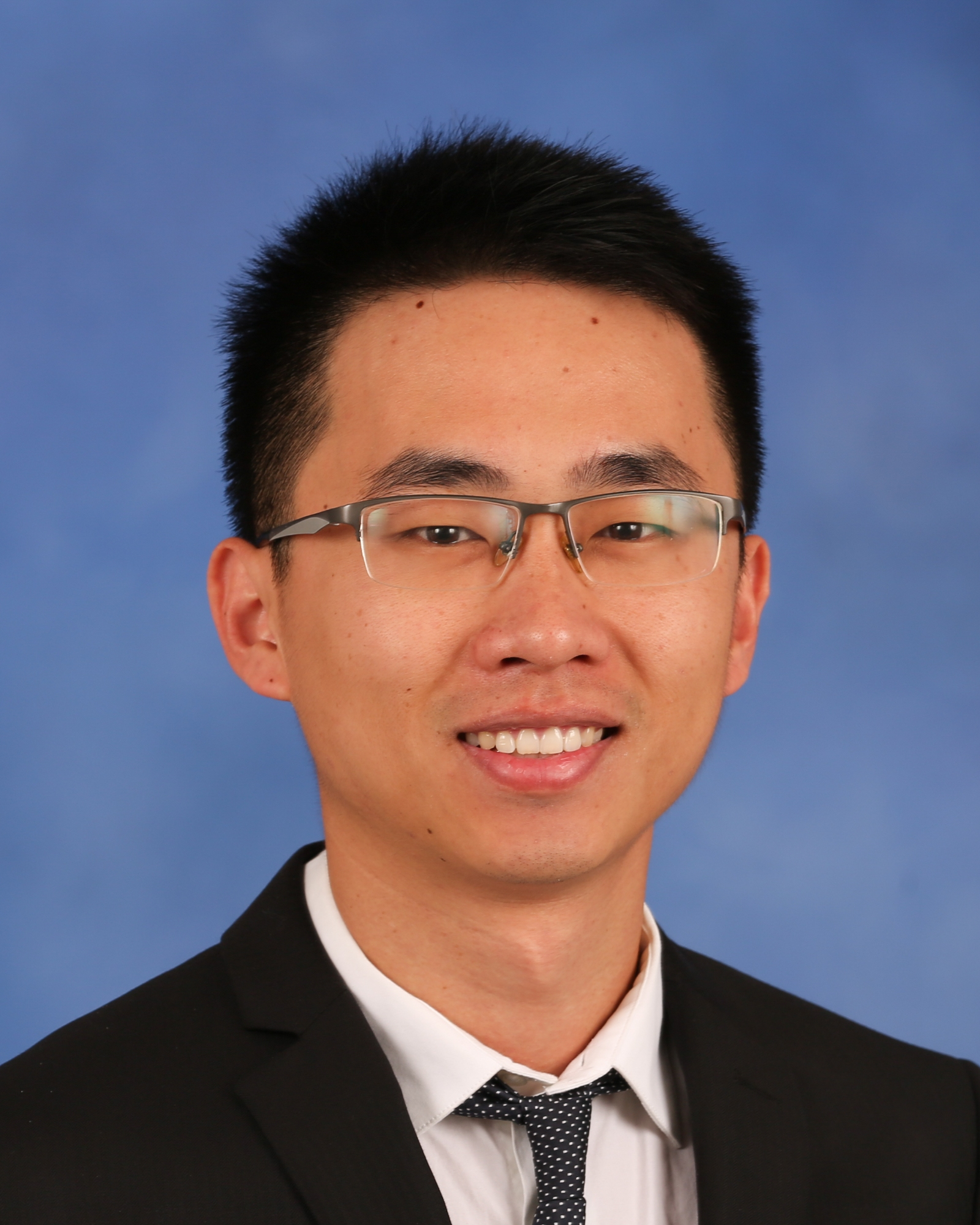}}]{Jianqing Liu} (M'18) received the Ph.D. degree from University of Florida in 2018. He is currently an assistant professor with the Department of Computer Science at NC State University. His research interest is wireless networking, security, and privacy. He received the U.S. National Science Foundation Career Award in 2021. He is also the recipient of several best paper awards including the 2018 Best Journal Paper Award from IEEE TCGCC.
\end{IEEEbiography}
\vspace{-.3in}
\begin{IEEEbiography}
[{\includegraphics[width=1in,height=1.25in,clip,keepaspectratio]{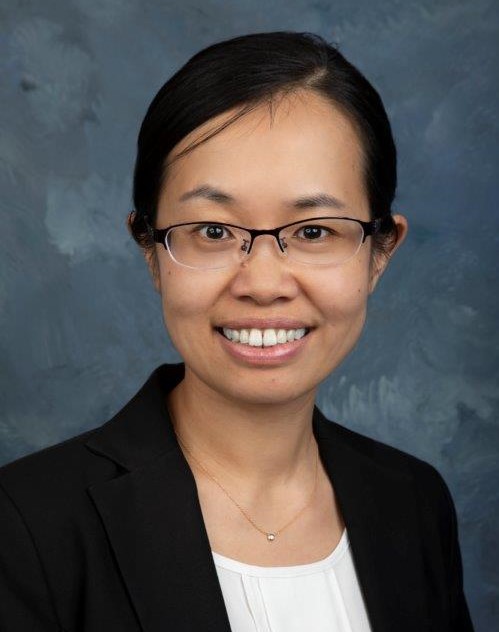}}]{Na Gong} (M’13) received the Ph.D. degree in computer science and engineering from the State University of New York, Buffalo, in 2013. Currently, Dr. Gong is a professor with the Department of Electrical and Computer Engineering at University of South Alabama. Her research interests include power-efficient computing circuits and systems, memory optimization, AI hardware, and neuromorphic computing. She is the recipient of the best paper nomination from ISVLSI’19, best paper award from EIT’16, best paper nominations from ISQED’16 and ISLPED’16. 
\end{IEEEbiography}
\vspace{-.3in}
\begin{IEEEbiography}
[{\includegraphics[width=1in,height=1.25in,clip,keepaspectratio]{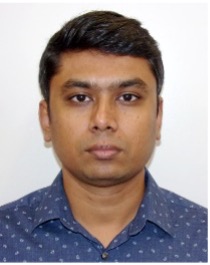}}]{Hritom Das} (M’20) received the Ph.D. degree in Electrical and Computer Engineering from North Dakota State University in 2020. He was a visiting Assistant Professor with the Department of Electrical and Computer Engineering at the University of South Alabama. Currently, he is a Post-Doctoral Research Associate with the Department of Electrical Engineering and Computer Science at The University of Tennessee. His research interests include neuromorphic computing and data privacy for edge devices.
\end{IEEEbiography}

\subsection*{Appendix A}
Recall that $\Delta A$ is defined as $O-X$, which measures how far apart the \texttt{SRAM\_DP} perturbed output $O$ is from the original input $X$. The closed form PMF of $\Delta A$ is unfortunately intractable because calculating $P(\Delta A)$ requires enumerating all combinations of $\left\{\Delta a_{n-1}, \ldots, \Delta a_{0}\right\}$ in an equality constraint (i.e., $\Delta A$ equals to a specific value) which is similar to the bounded knapsack problem that is NP-complete. This poses a great challenge to analyze the $l_1$ utility loss of \texttt{SRAM\_DP} in Theorem \ref{DP2}. In light of it, this section will not seek to derive the closed form PMF of $\Delta A$, but rather to prove that the PMF of $\Delta A$ has a zero mean and is symmetric w.r.t. $\Delta A = 0$. This key property will facilitate us to arrive at the conclusion in Theorem \ref{DP2}.

First, for notation simplicity, we introduce a new variable $S_{n}(a)$ to denote the probability of $\Delta A = a$ for any $n$-bit data ($n \ge 1$), i.e., $S_{n}(a)=P(\Delta A=a)$. Note that $S_{n}(a)$ has two key properties. (i) $a$ is defined as $a \in \mathbb{Z}$, but $S_{n}(a) = 0$ for $a \notin \{-2^{n}+1,...,2^{n}-1\}$; (ii) $S_{n}(a)$ has a following boundary condition concurring with Eq.(\ref{pmf}) 
\[ S_{1}(a) = \begin{cases}
      \frac{1}{4}f_1 & a = -1 \\
      1-\frac{1}{2}f_1 & a = 0\\
      \frac{1}{4}f_1 & a = 1 
   \end{cases}
\]

Next, we develop a recursive approach to prove Lemma \ref{DP2_lemma}. First of all, $S_{n}(a)$ has the following the recursion relation
\begin{equation}\label{recursion}
\begin{aligned}
S_{n}(a)= & \frac{1}{4} f_{n} S_{n-1}\left(a-2^{n-1}\right)+\left(1-\frac{1}{2}f_{n}\right) S_{n-1}(a) \\
&+\frac{1}{4} f_{n} S_{n-1}\left(a+2^{n-1}\right).
\end{aligned}
\end{equation}
The intuition behind Eq.(\ref{recursion}) is that a $n$-bit data can have $\Delta A = a$ when all bits except the $n^{\text{th}}$ bit are (i) equal to $(a-2^{n-1})$ in decimal while the $n^{\text{th}}$ bit fails (i.e., $\Delta a_n = 1$); or (ii) equal to $a$ in decimal while the $n^{\text{th}}$ bit is intact (i.e., $\Delta a_n = 0$); or (3) equal to $(a+2^{n-1})$ in decimal while the $n^{\text{th}}$ bit fails (i.e., $\Delta a_n = -1$).

Then, we follow the principle of \emph{proof by induction} to prove Lemma \ref{DP2}. The standard proof procedures are as follows. (i) From the the boundary condition, we know that Lemma \ref{DP2} stands true for the base case $S_{1}(a)$. (ii) Assume Lemma \ref{DP2} also holds for $S_{k}(a)$ where $2 \leq k < n$. This implies that $S_{k}(a)$ = $S_{k}(-a)$, $S_{k}(a - 2^k)$ = $S_{k}(2^k-a)$, and $S_{k}(-a - 2^k)$ = $S_{k}(a+2^k)$. Then, according to Eq.(\ref{recursion}), we can derive
\begin{equation*}
\begin{aligned}
S_{k+1}(-a)= & \frac{1}{4} f_{k+1} S_{k}\left(-a-2^{k}\right)+\left(1-\frac{1}{2}f_{k+1}\right) S_{k}(-a) \\
&+ \frac{1}{4} f_{k+1} S_{k}\left(-a+2^{k}\right) \\
= & S_{k+1}(a).
\end{aligned}
\end{equation*}
Therefore, by the principle of \emph{proof by induction}, we can assert that $S_{n}(a)$ is symmetric w.r.t. to $a = 0$, $\forall n \ge 1$ (naturally, it has a zero mean). In other words, the PMF of $\Delta A$ also obeys the same property, which concludes the correctness of Lemma \ref{DP2}.

\subsection*{Appendix B}
\subsubsection*{B.1. Voltage Control}
\texttt{SRAM\_DP} adopts four 6T cells to store LSB1-4 in order to inject noise. Its failure rate under different supply voltages is shown in Table \ref{tab:fail:dp}. The corresponding $\epsilon$ at any specific supply voltage can be easily calculated according to Theorem \ref{DP}. In differential privacy, $\epsilon$ is a rather abstract quantity and recent research articles can select $\epsilon$ from as little as 0.01 to as much as 7 \cite{hsu2014differential}. Our proposed memory can also achieve a wide-range adaptation of $\epsilon$ under broad tuning of supply voltages. Yet, the chip stability can be hardly controlled at extreme low-voltage operations (i.e., when cell failure rate exceeds 90\%). Therefore, we only showcase the results in Table \ref{tab:fail:dp}. 

Based on the failure rates listed in Table \ref{tab:fail:dp}, the following four permutation patterns from MSB to LSB are considered for \textbf{Bit Shift} in Step 1: $\pi_{1}$-[0, 1, 2, 3, 4, 5, 6, 7]; $\pi_{2}$-[0, 1, 2, 3, 5, 4, 7, 6]; $\pi_{3}$-[0, 1, 2, 3, 6, 7, 4, 5]; and $\pi_{4}$-[0, 1, 2, 3, 7, 6, 5, 4]. Accordingly, during the data storage process, two bits are sufficient for permutation pattern selection and they are stored in memory as shown in Figure~\ref{fig:memory_overview}.
\begin{table}
\caption{Cell Failure Rates and Corresponding $\epsilon$'s}
\centering
 \begin{tabular}{c | c | c} 
 \hline
 Supply Voltage (V) & Failure Rate (\%) & $\epsilon$  \\ 
 \hline\hline
 0.50 & 81.57 & 1.49  \\ 
 \hline
 0.55 & 70.57 & 2.43  \\
 \hline
 0.56 & 68.31 & 2.63  \\
 \hline
 0.57 & 66.15 & 2.82  \\
 \hline
 0.58 & 64.09 & 3.01 \\ 
 \hline
 0.59 & 62.03 & 3.20  \\
 \hline
 0.60 & 60.26 & 3.36 \\ 
 \hline
\end{tabular}
\label{tab:fail:dp}
\end{table}

\subsubsection*{B.2. Custom Cells Design} 
\begin{figure}
\centering
 \includegraphics[width=\linewidth]{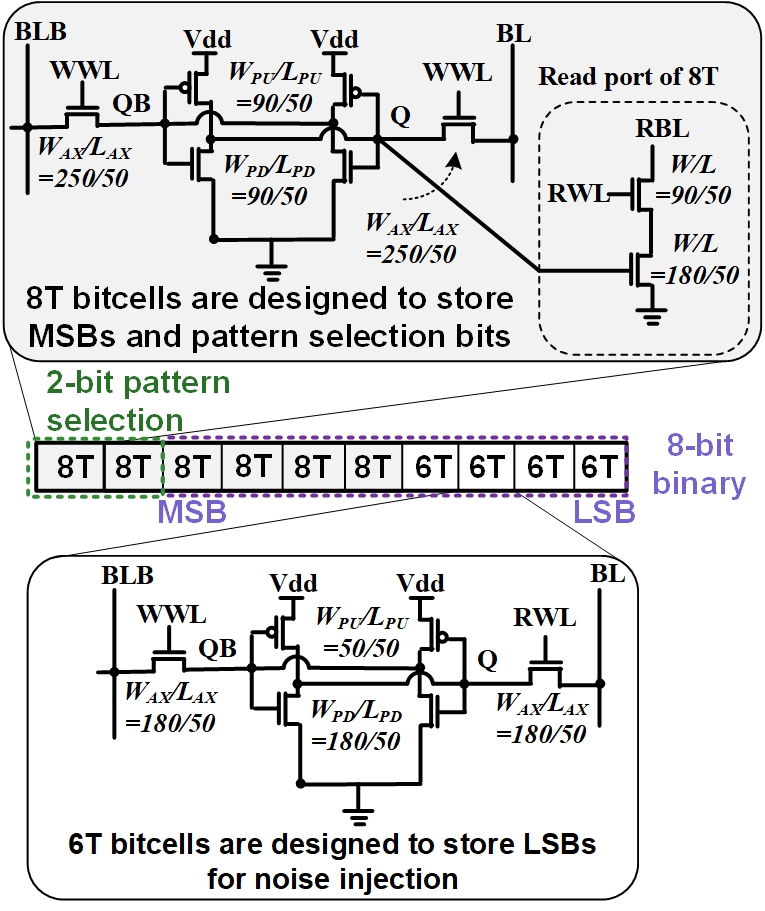}
  \caption{Cell design for the proposed memory.}
  \label{fig:cell}
\end{figure}

As discussed earlier, to implement the proposed \texttt{SRAM\_DP}, we need to render cell failures to the four LSBs (i.e., LSB1-4), while maintaining four MSBs (i.e., LSB5-8) intact. To enable it, we propose a hybrid-cell design scheme, as shown in Figure~\ref{fig:cell}. Reliable 8T cell are custom designed to enable zero failure rate and they are used to store two pattern selection bits and four MSBs of the input data (i.e., LSB5-8)) to minimize the utility loss. Area-efficient 6T cells are custom designed to store the four LSBs (i.e., LSB1-4). To achieve zero area overhead of 6T–8T integration, two separate word lines (RWL and WWL) are used for 6T cells \cite{Gong2012Cell}.

\subsubsection*{B.3. Bit Shuffler and Re-Shuffler Design} 
\begin{figure}
\centering
  \includegraphics[width=\linewidth]{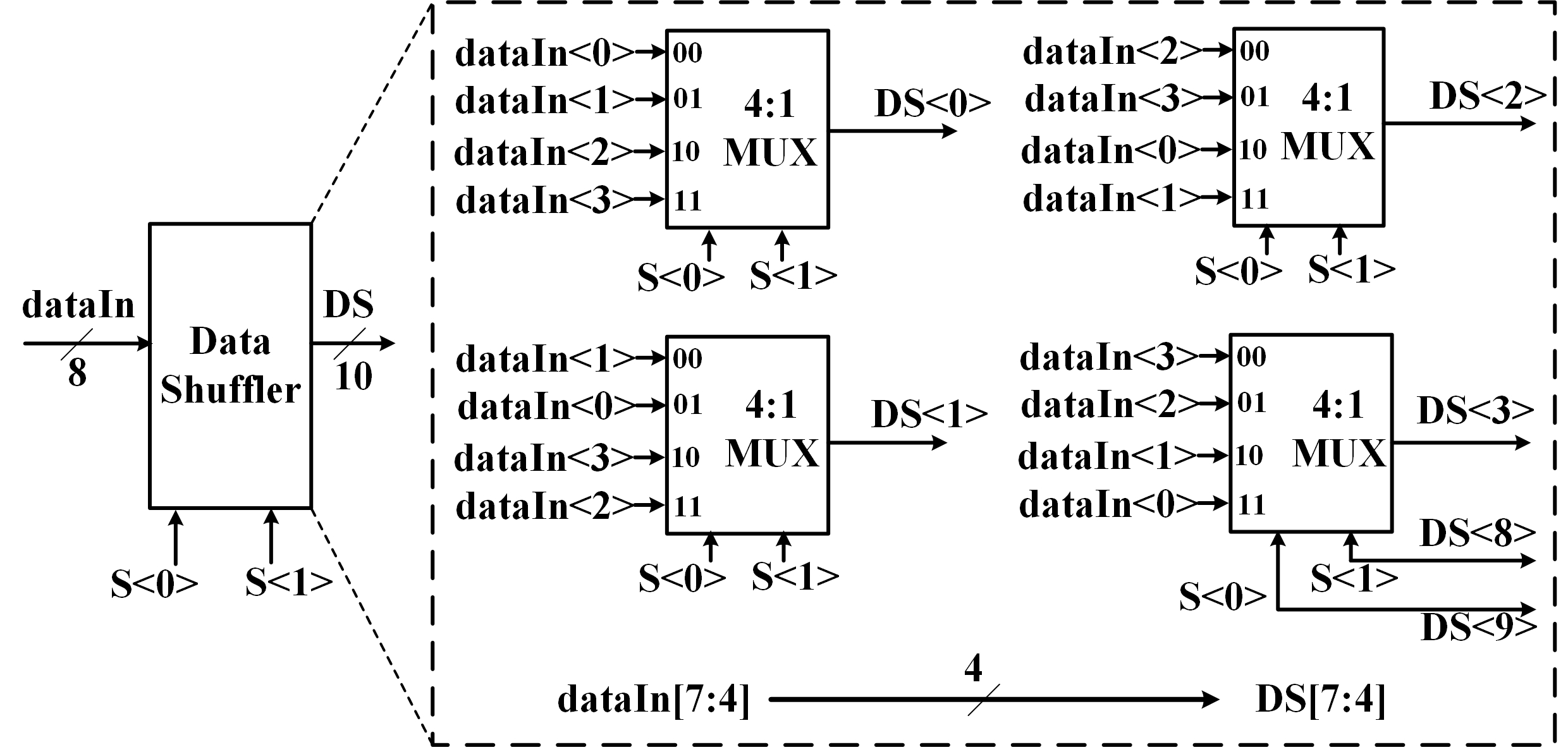}
  \caption{Bit shuffler.}
  \label{fig:data_shuffler}
\end{figure}

A MUX-based bit Shuffler is designed to enable \textbf{Bit Shift} in Step 1, as shown in Figure~\ref{fig:data_shuffler}. The 8-bit input data called dataIn will be shuffled according to a 2-bit pattern selection signal S[1:0]. Specifically, four 4-to-1 MUXs are adopted and each MUX is used to shift one LSB. As S[1:0] is ``00'', ``01'', ``10'', or ``11'', $\pi_{1}$ to $\pi_{4}$ will be selected accordingly. For example, if S is ``11'', dataIn$\langle 3 \rangle$ is selected as DS[0], dataIn$\langle 2 \rangle$ is selected as DS[1], dataIn$\langle 1 \rangle$ is selected as DS[2], and dataIn$\langle 0 \rangle$ is selected as DS[3]. The four MSBs of the output DS will remain the same as dataIn. As a result, $\pi_{4}$-[0, 1, 2, 3, 7, 6, 5, 4] is selected for shuffling. During this process, a pattern selection signal S is generated by a 2-bit random number generator, which will be stored in the same memory word (row) together with the shuffled data for the \textbf{Reverse Bit Shift} in Step 4. Accordingly, the output of the bit shuffler DS has 10 bits, including 8-bit shuffled data and 2-bit pattern selection signal S, as shown in Figure~\ref{fig:data_shuffler}.  

To avoid performance penalty, the Bit Shuffler is operated in parallel with the word-line decoder of the memory. Due to its small size, the access delay of Bit Shuffler is quite marginal and can be overlapped with the word-line decoding. Therefore, the Bit Shuffler will not create new critical paths in the memory. At the same time, the Bit Shuffler also introduces a small amount of power overhead. However, the power savings achieved by the proposed low-voltage memory can offset this overhead. 

\begin{figure}
\centering
  \includegraphics[width=\linewidth]{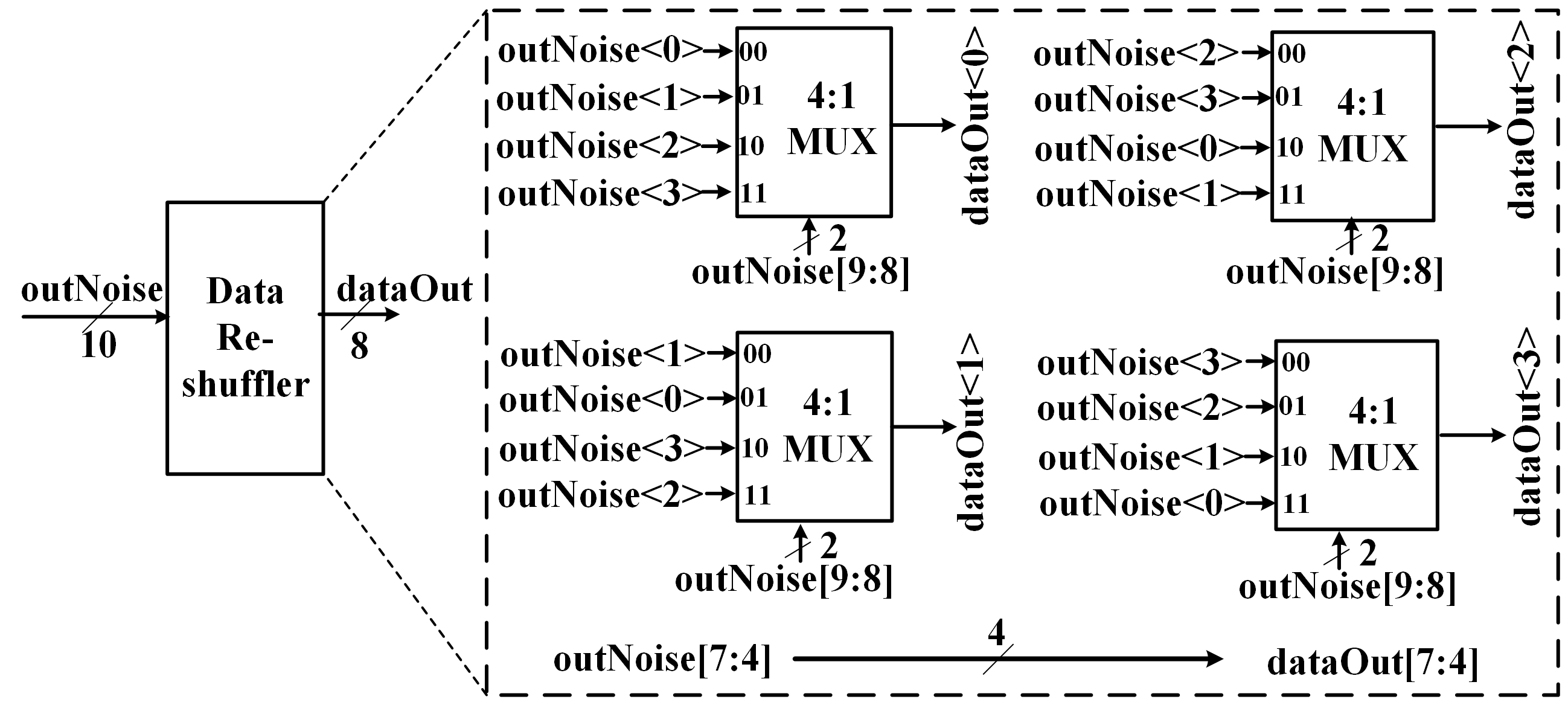}
  \caption{Bit reshuffler.}
  \label{fig:data_reshuffler}
\end{figure}

A MUX-based Bit Re-Shuffler is also designed to implement the \textbf{Reverse Bit Shift} in Step 4, as shown in Figure~\ref{fig:data_reshuffler}. The loaded two pattern selection bits (outNoise$\langle 9 \rangle$ and outNoise$\langle 8 \rangle$, i.e., S$\langle 1 \rangle$ and S$\langle 0 \rangle$) are used to revert the memory output data outNoise[7:0] according to the original bit orders. It is worthwhile to emphasize that The proposed MUX-based Shuffler and Re-Shuffler schemes is only dependent on the permutation vector set, and it can easily adapt to different permutation patterns, with minimal implementation cost.

\subsubsection*{B.4. Random Bit Generator} 
\begin{figure}
\centering
  \includegraphics[width=\linewidth]{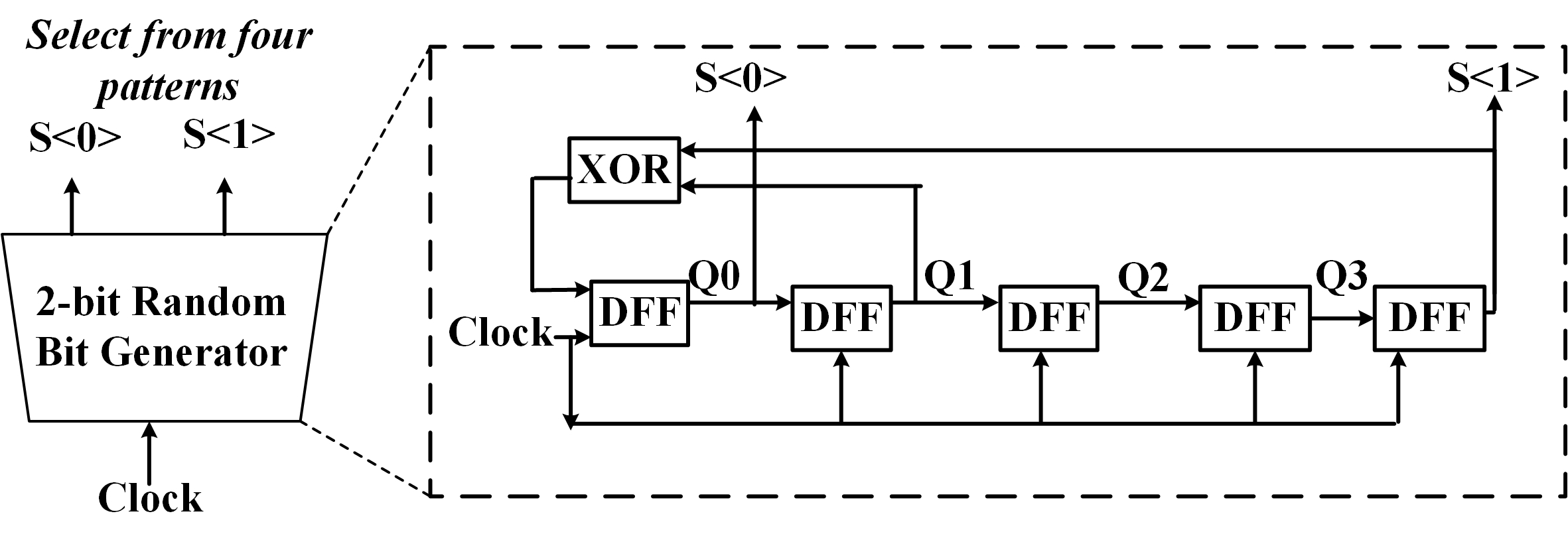}
  \caption{2-bit random bit generator for pattern selection.}
  \label{fig:random}
\end{figure}

This module is needed in the \textbf{Step 1: Bit Shift} to randomly select one permutation pattern, and in the \textbf{Step 3: Noise Injection} to add random noise to the failed positions of the memory. In this paper, we adopt a light-weight Random Bit Generator design using Linear Feedback Shift Registers (LFSR), as shown in Figure~\ref{fig:random}. To improve the randomness, a true random generator, such as \cite{clark2018sram}, can be used, which, however, will come with a significant implementation cost. 

\end{document}